\documentclass[aip,amsmath,amssymb,preprint]{revtex4-1}

\usepackage{graphicx,amssymb,bm,amsfonts,amsmath,graphics,color,epstopdf}
\usepackage[bookmarks=false,pdfstartview=FitH,colorlinks=true,citecolor=blue,linkcolor=blue]{hyperref}
\usepackage{lineno}
\usepackage{physics}
\usepackage{float}
\begin{document}

\title{Polarization dependent photoemission as a probe of the magnetic ground state in the layered ferromagnet VI$_3$}

\author{Derek Bergner} 
\affiliation{Department of Physics and Astronomy, California State University Long Beach, Long Beach, California 90840, USA}

\author{Tai Kong}
\thanks{Current affiliation: Department of Physics, University of Arizona, Tucson Arizona, 85721, USA}
\affiliation{Department of Chemistry, Princeton University, Princeton, New Jersey, 08544, USA}

\author{Ping Ai}
\affiliation{Department of Physics, University of California Berkeley, California 94720, USA}
\affiliation{Materials Sciences Division, Lawrence Berkeley National Laboratory, Berkeley, California 94720, United States}

\author{Daniel Eilbott}
\affiliation{Department of Physics, 
University of California Berkeley, California 94720, USA}
\affiliation{Materials Sciences Division, Lawrence Berkeley National Laboratory, Berkeley, California 94720, United States}

\author{Claudia Fatuzzo}
\affiliation{Department of Physics, University of California Berkeley, California 94720, USA}
\affiliation{Materials Sciences Division, Lawrence Berkeley National Laboratory, Berkeley, California 94720, United States}

\author{Samuel Ciocys}
\affiliation{Department of Physics, University of California Berkeley, California 94720, USA}
\affiliation{Materials Sciences Division, Lawrence Berkeley National Laboratory, Berkeley, California 94720, United States}

\author{Nicholas Dale}
\affiliation{Department of Physics, University of California Berkeley, California 94720, USA}
\affiliation{Materials Sciences Division, Lawrence Berkeley National Laboratory, Berkeley, California 94720, United States}

\author{Conrad Stansbury}
\affiliation{Department of Physics, University of California Berkeley, California 94720, USA}
\affiliation{Materials Sciences Division, Lawrence Berkeley National Laboratory, Berkeley, California 94720, United States}

\author{Drew Latzke}
\affiliation{Department of Physics, University of California Berkeley, California 94720, USA}
\affiliation{Materials Sciences Division, Lawrence Berkeley National Laboratory, Berkeley, California 94720, United States}

\author{Everardo Molina}
\affiliation{Department of Physics and Astronomy, California State University Long Beach, Long Beach, California 90840, USA}

\author{Ryan Reno}
\affiliation{Department of Physics and Astronomy, California State University Long Beach, Long Beach, California 90840, USA}

\author{Robert J. Cava}
\affiliation{Department of Chemistry, Princeton University, Princeton, New Jersey, 08544, USA}

\author{Alessandra Lanzara}
\affiliation{Department of Physics, University of California Berkeley, California 94720, USA}
\affiliation{Materials Sciences Division, Lawrence Berkeley National Laboratory, Berkeley, California 94720, United States}

\author{Claudia Ojeda-Aristizabal}
\affiliation{Department of Physics and Astronomy, California State University Long Beach, Long Beach, California 90840, USA}

\date{\today}

\begin{abstract}
Layered ferromagnets are thrilling materials from both a fundamental and technological point of view. VI$_3$ is an interesting example, with a complex magnetism that differentiates it from the first reported Cr based layered ferromagnets. Here, we show in an indirect way through Angle Resolved Photoemission Spectroscopy (ARPES) experiments, the importance of spin-orbit coupling setting the electronic properties of this material. Our light polarized photoemission measurements point to a ground state with a half-filled $e_\pm'$ doublet, where a gap opening is triggered by spin-orbit coupling enhanced by electronic correlations. 
\end{abstract}

\pacs{}

\maketitle

Layered ferromagnets, van der Waals materials that exhibit a magnetic order at the level of single layer, bring a new functionality to the family of 2-dimensional crystals, opening the door to ultrathin spintronic applications. Vanadium(III) iodide ($VI_3$) is a van der Waals (vdW) Mott insulator made of honeycomb vanadium layers separated by an iodine-iodine vdW gap with the remarkable property of intrinsic ferromagnetism. Despite the fact that VI$_3$ has proved to be extremely sensitive to air, its recently reported unusual thickness-dependent magnetism with an increase of the Curie temperature as the layer number decreases \cite{doi:10.1021/acs.nanolett.1c03027}, makes it a compelling platform for the engineering of magnetism in electronic devices. Angle Resolved Photoemission Spectroscopy (ARPES) is an excellent tool to study the electronic properties of layered materials. Its ultrahigh vacuum environment and in-situ cleaving capability reduces the limitations and challenges associated to probing air-sensitive materials.    

\begin{figure}[H]
\includegraphics[width=7cm]{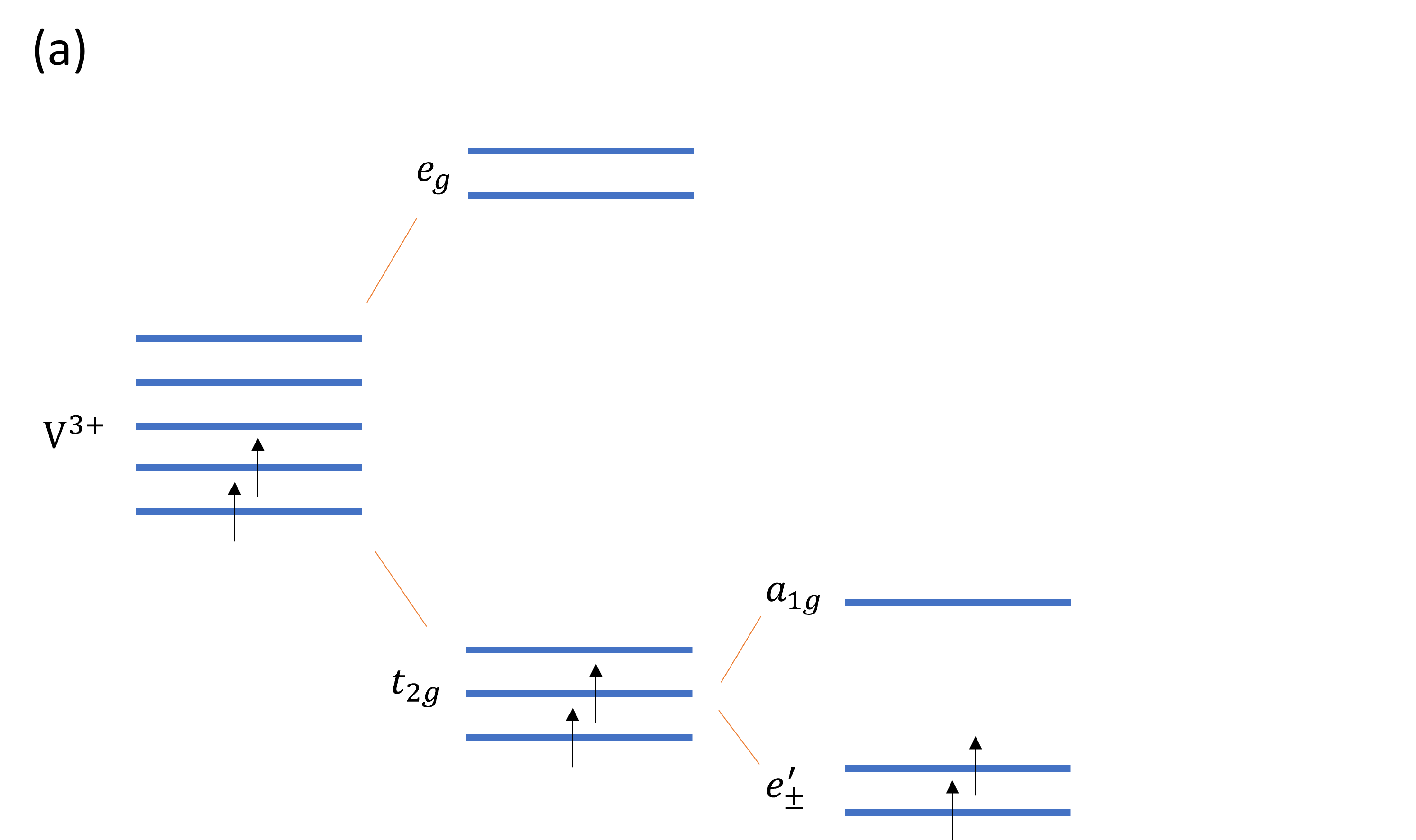}
\includegraphics[width=7.5cm]{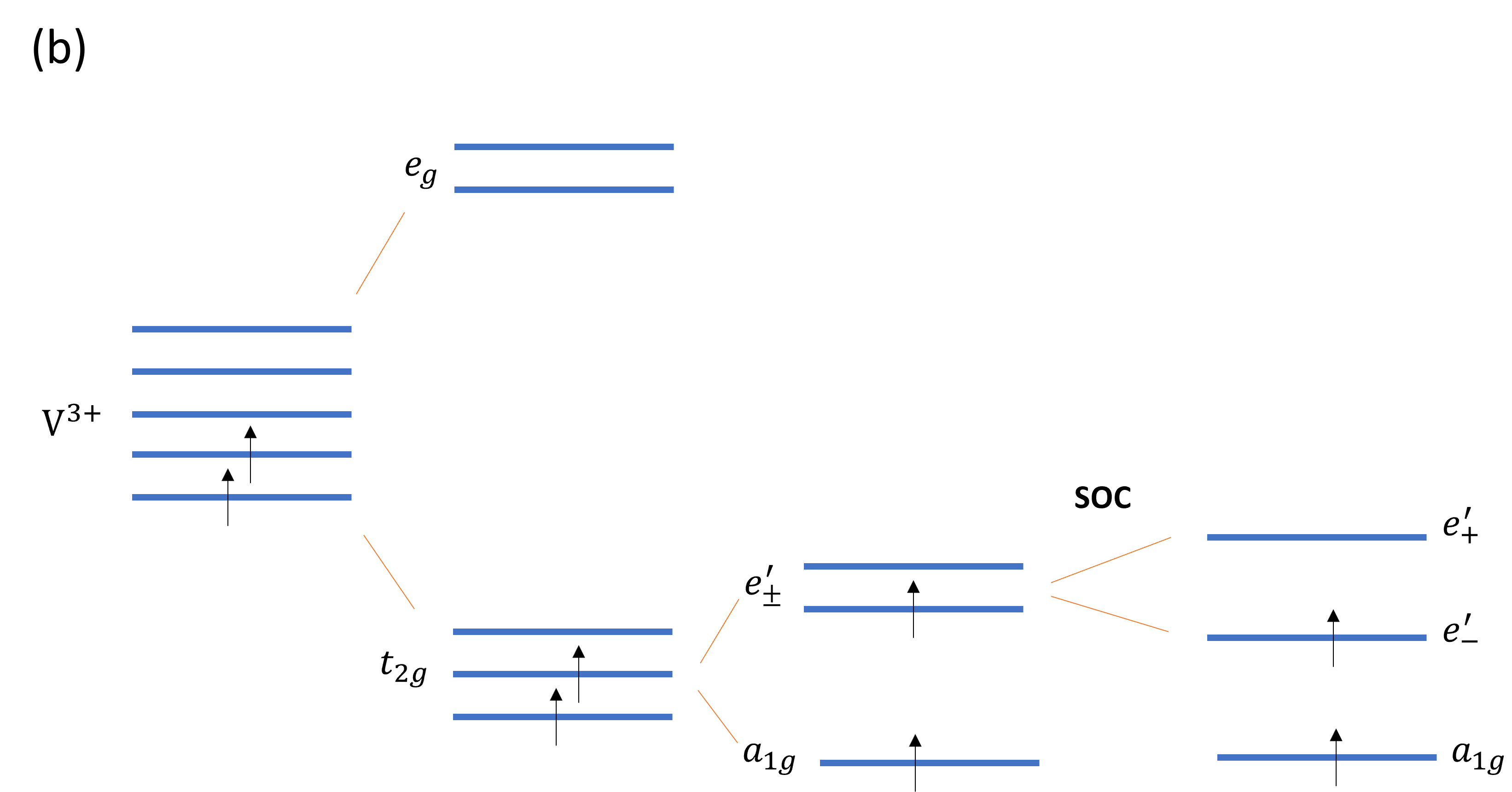}
\caption{\label{EnergyLevels} \textbf{Splitting of the d energy levels in V 3d$^2$} in the $e_{\pm}^{2}a_{1g}^0$ (a) and $a_{1g}^1e_{\pm}^{1}$ (b) configurations, the last one with active SOC coupling} 
\end{figure}

VI$_3$'s electronic properties derive from a combination of crystal field effects, electronic correlations and spin-orbit coupling. In VI$_3$, vanadium ions (V) are surrounded by I$_6$ octahedra which provides a crystal field that splits the five-fold degenerate V-d orbitals into a lower triplet t$_{2g}$ and an upper doublet e$_g$. Additional trigonal distorsions caused by the elongation of of VI$_6$ octahedra along one of the [111] axes \cite{khomskii_2014}, causes an additional split of the t$_{2g}$ levels into a singlet a$_{1g}$ and a doublet e$_\pm'$ (see Figure \ref{EnergyLevels}). $|a_{1g}\rangle$ has a simple form in coordinates where the z axis aligns with the trigonal distorsion, and x and y axes are in the perpendicular plane, pointing along the z-direction. The other two t$_{2g}$ orbitals $|e_\pm'\rangle$ are complex and have the shape of a torus \cite{khomskii_2014, PhysRevB.104.014414}. 

There is not yet consensus on what is the exact mechanism responsible of the observed bandgap in VI$_3$. The two V d$^2$ electrons in VI$_3$ may occupy the $|a_{1g}\rangle$ and $|e_\pm'\rangle$ orbitals into the state $e_{\pm}^{2}a_{1g}^0$, that together with electronic correlations leads to a Mott insuator \cite{PhysRevB.104.014414, PhysRevB.101.024411, C9CP05643B}, or electrons may partially fill the $|e_\pm'\rangle$ orbitals, which leads in principle to a metallic ground state $a_{1g}^1e_{\pm}^{1}$; however spin-orbit coupling 
is capable of spliting the half filled $|e_\pm'\rangle$, that in conjunction with electronic correlations leads also to a Mott insulator \cite{PhysRevB.101.100402, C9CP05643B}.  Both scenarios have been proved possible through first principle calculations, and are consistent with the known S=1 state for VI$_3$ \cite{PhysRevB.101.100402, C9CP05643B}.

Understanding the t$_{2g}$ orbitals in VI$_3$ is fundamental, as they not only set the insulating character of this vdW material but also determine its magnetic properties. Perpendicular magnetic anisotropy, that refers to the preferred orientation of magnetic vectors perpendicular to the layer plane, is a fundamental characteristic in magnetic materials for applications, and most importantly, it is essential for the existence of magnetic order at low dimensions \cite{MerminWagnerFerro}. Perpendicular magnetic anisotropy comes from a microscopic effect, called single ion anisotropy or magnetocrystalline anisotropy, caused by the orbital momentum along the normal of the bonding plane quenching due to crystalline fields. The first layered ferromagnets reported in the literature were Cr based \cite{Huang2017,Gong2017}, such as CrI$_3$, where crystal field effects also split the 5-fold degenerate Cr-d orbitals into t$_{2g}$ and e$_g$ states however, the degeneracy of the t$_{2g}$ orbital for Cr d$^{3}$ is unlifted 
and yields no magnetocrystalline anisotropy. Perpendicular magnetic anisotropy in CrI$_3$ comes rather from the weak exchange anisotropy brought by the heavy I 5p orbitals and their hybridization with Cr 3d orbitals \cite{PhysRevB.101.100402,Lado_2017,PhysRevLett.122.207201}. 
VI$_3$ has a large magnetocrystalline anisotropy (one order of magnitude larger than for CrI$_3$) that together with spin-orbit coupling, is responsible for the observed perpendicular magnetic anisotropy. In VI$_3$, it is spin-orbit coupling that aligns the magnetic moment of the V ions along the z-axis via the strong magnetocrystalline anisotropy and produces the observed perpendicular magnetic anisotropy\cite{PhysRevB.101.100402}. Angle-dependent magnetization measurements 
have demonstrated that the magnetic moment in VI$_3$ is in fact canted by $\approx$40$^{\circ}$ from the normal to the ab plane in a wide temperature range (10 K - 60 K), within the range of the two ferromagnetic phases reported for VI$_3$ as well as its structural phase transition. The observed magnetization canted easy-axis differentiates VI$_3$ from the Ising-type magnetism reported for CrI$_3$ \cite{PhysRevB.103.174401}.

 The orientation of the V 3d orbitals that host the magnetic Mott ground state, discernible through the use of different light polarizations in ARPES, can give insight on the ground state that brings magnetism to this vdW material.  
Previous photoemission experiments have characterized the valence band structure of bulk VI$_3$ \cite{Kundu2020} however, further understanding of the character of the observed band manifolds remains elusive. 
Here we report a deep study on the linear and circular dichroism of VI$_3$ photoemission, that together with the effects of light energy and alkaline doping brings insight on the orbitals' properties that play an important role in 2-dimensional magnetism.

VI$_3$ millimeter-sized single crystals were synthesized by a chemical vapor transport method as reported in \onlinecite{TaiKong}. A stoichiometric mixture of vanadium powder and iodine pieces was loaded in a silica tube and sealed under vacuum. Growth was performed over 3-4 days in a horizontal tube furnace, with the hot end held at 400 °C and the cold end at ambient temperature. High-resolution ARPES experiments were performed at Beamline 4.0.3 (MERLIN) of the Advanced Light Source using 30 - 128 eV linearly or circularly polarized photons in a vacuum better than 5 × 10$^{-11}$ Torr. The total-energy resolution was 20 meV with an angular resolution $\Delta\theta$ of $\leq$0.2\textdegree. 

\begin{figure}[H] 
\includegraphics[width=16cm]{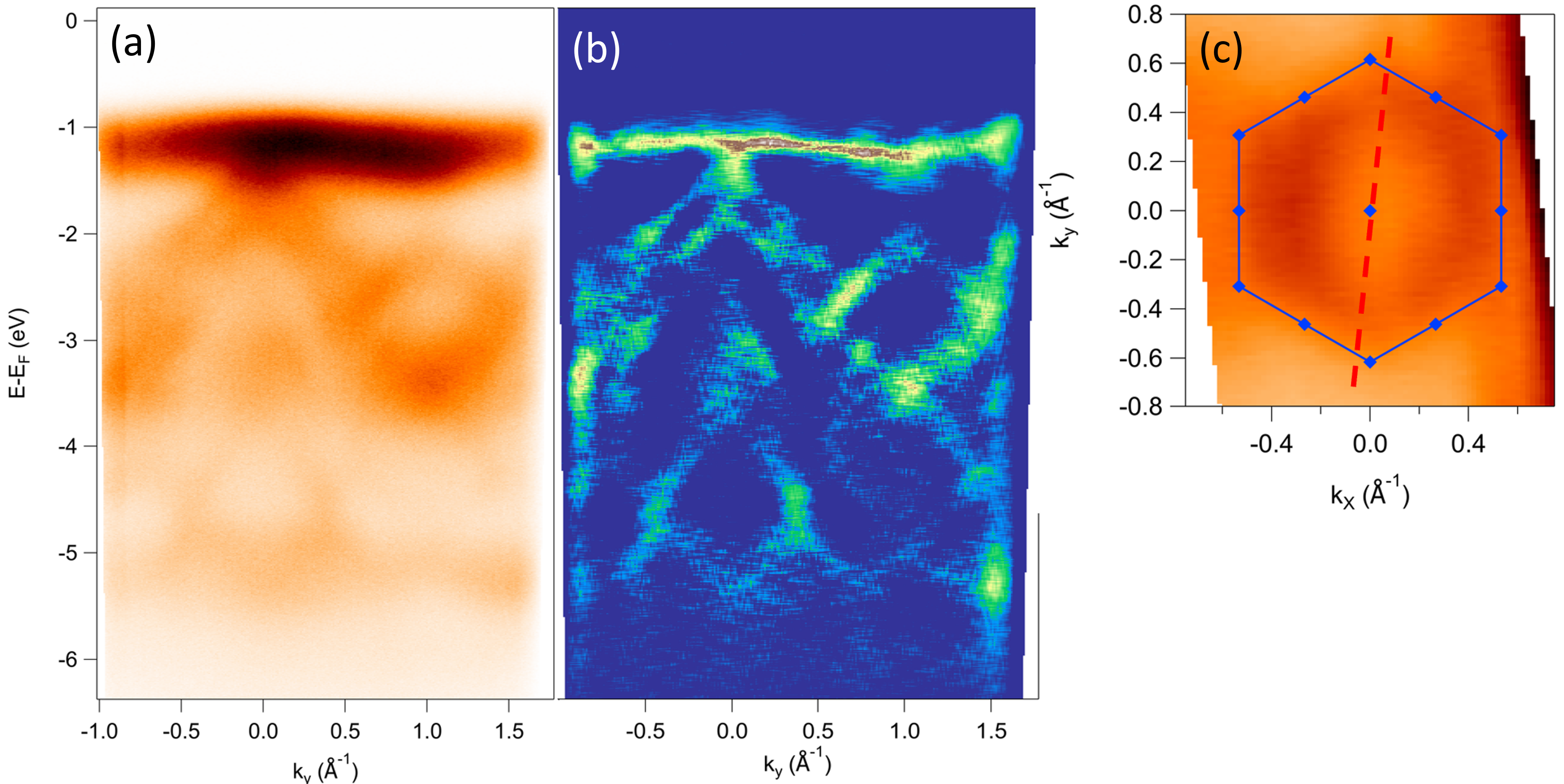}\caption{\label{Figure1} \textbf{VI$_3$ band structure}  a) ARPES measured along a line close to the high symmetry direction K-$\Gamma$-K, as indicated by the pointed line in  (c). (b) Curvature of the measured electronic band structure shown in (a). (c) Constant energy map taken at 4.5 eV below the Fermi energy. The blue hexagon indicates the first Brillouin zone.} 
\end{figure}

Figure \ref{Figure1} shows the measured electronic band structure at T$=264$ K using photon energy h$\nu=76$ eV and out-of-plane linearly polarized light, within the setup in Figure \ref{Figure2}e (detailed later). Measurements were done at 265 K to avoid charging of the sample. Despite the fact that this is well above the magnetic transition temperature reported for VI$_3$ ($\approx$50 K\cite{TaiKong}), previous ARPES measurements report no major differences in the band structure at 35 K and 300 K \cite{Kundu2020}. We calculate the curvature of the data in the energy direction (Figure \ref{Figure1} (a)), using the method elaborated in \onlinecite{doi:10.1063/1.3585113} . At the high symmetry direction K-$\Gamma$-K, we observe a non-dispersive band at the top of the conduction band ($\approx$ 1eV below the Fermi level) and dispersive bands at higher binding energies.  
A constant energy map at 4.5 eV below the Fermi energy (\ref{Figure1}(c)), reveals a six-fold symmetry, consistent with the known trigonal structure for VI$_3$ at room temperature. The superimposed blue hexagon corresponds to the first Brillouin zone, calculated using the parameters a=b=6.89 \AA$^{‐1}$ and c=19.81 \AA$^{‐1}$ reported in the literature from single crystal x-ray diffraction experiments \cite{TaiKong, PhysRevMaterials.3.121401}. As in reference \onlinecite{Kundu2020}, we identify the flat band near the top of the valence band of V 3d character. As stated by Kundu et al, other V based vdW materials such as VSe$_2$\cite{VSe2_ACSNano} and VTe$_2$\cite{VTe2_PRB}, show equally weakly dispersing V 3d-electron-derived bands. The attributed V 3d character of this band at $\approx$1 eV below the Fermi energy is also consistent with our photon energy dependent photoemission data. Following the work by Kundu et al., we compare the calculated photoemission crosssections \cite{CrossSectionParry, YEH19851} to energy distribution curves (EDC) for different photon energies (See Supplementary Materials), where we verify that V 3d-electron derived bands show a stronger intensity for certain photon energies (vs I 5p bands) as listed in \onlinecite{CrossSectionParry} and \onlinecite{YEH19851}.  

\begin{figure}[H]
\includegraphics[width=15cm]{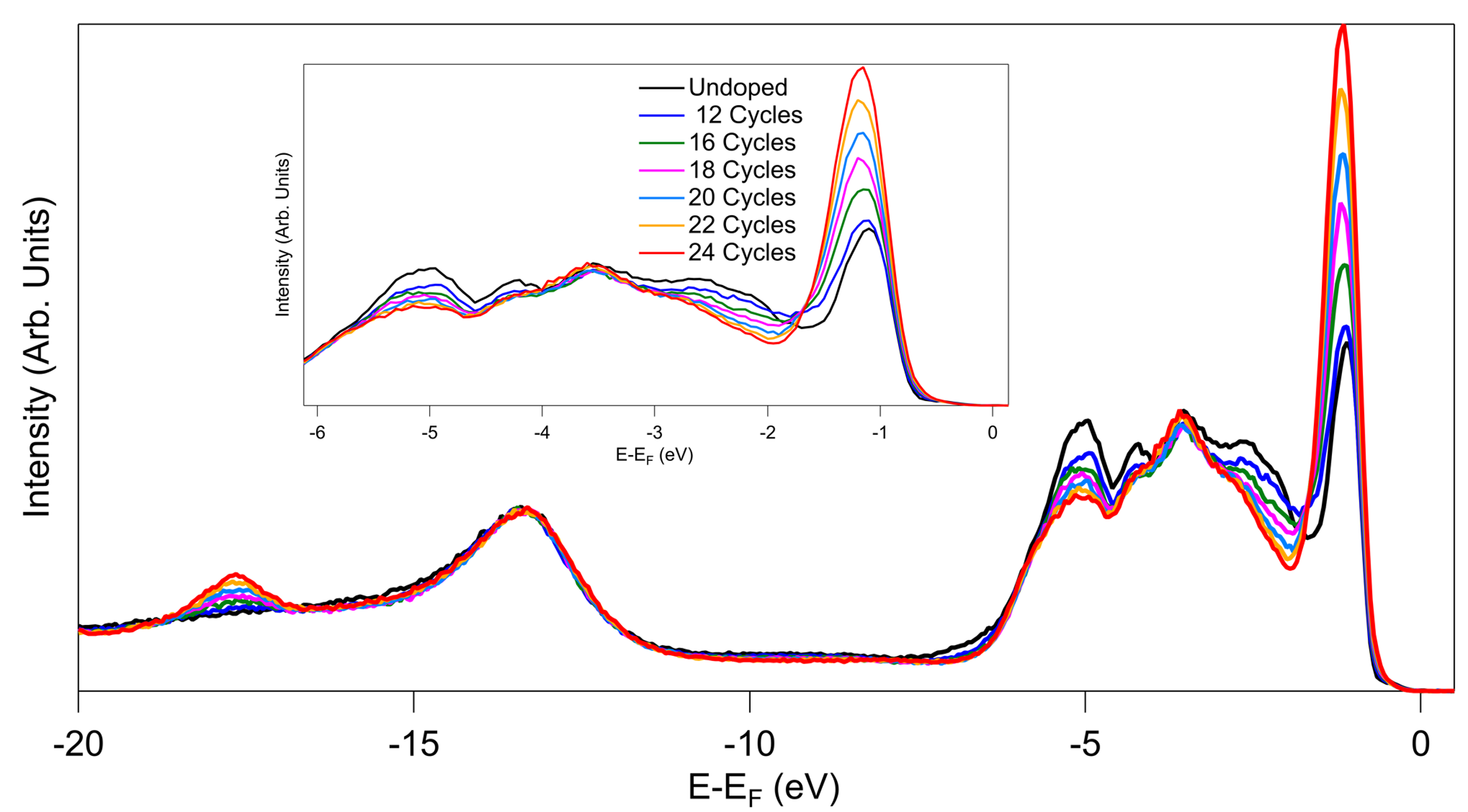}\caption{\label{FigureDop} \textbf{Effect of in-situ K doping}  Momentum-integrated energy distribution curves after multiple cycles of doping. Each cycle corresponds to 4.0 Amps applied to a K effusion cell during 1 minute. The red curve corresponds to no-doping, the dark blue curve is the highest K doping. The K associated peak (17.5 eV) grows through the doping cycles.} 
\end{figure}

\begin{figure}[H]
\includegraphics[width=17cm]{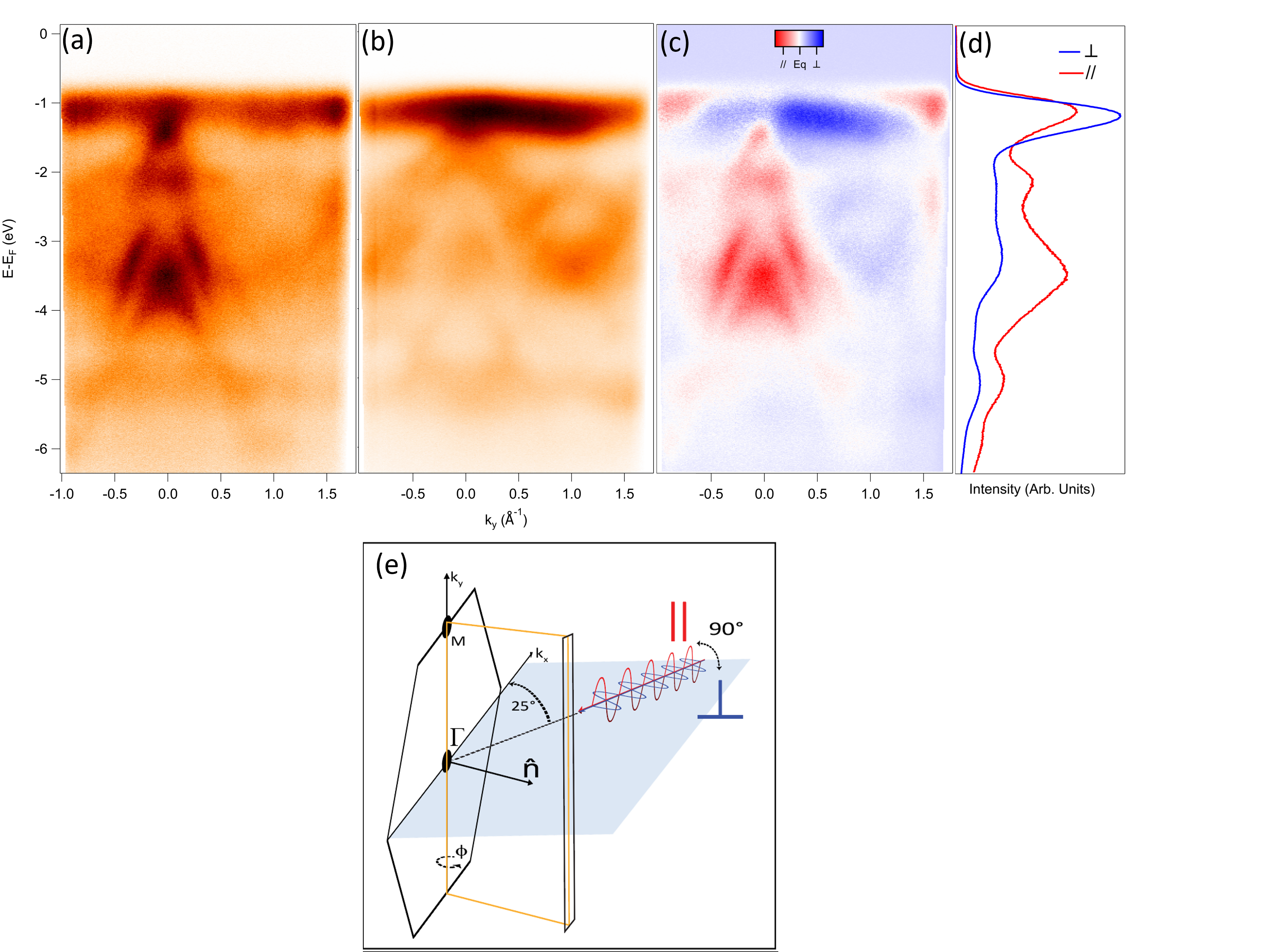}\caption{\label{Figure2}\textbf{Linear polarization dependent VI$_3$ band structure} using in-plane (a) and out of plane (b) light polarization. (c) Linear dichroism calculated from the difference in intensities in figures (a) and (b). (d) Momentum integrated Energy Distribution Curve mEDCs for out-of-plane light polarization (blue curve) and in-plane light polarization (red curve). (e) Schematics of the experimental setup showing incident photons polarized in-plane $\parallel$ (red) and out-of-plane $\perp$ (blue). Photons are incident on the blue plane 25$^{\circ}$ with respect to the surface of the sample (65$^{\circ}$ with respect to the normal of the sample $\hat{n}$). The rectangle in black represents the detector slit and in orange the detector plane. M and $\Gamma$ label high symmetry points of the first Brillouin zone and $\phi$ the polar angle that allows to reach different k$_x$. Data presented in the figure is taken at normal emission, $\phi=0$.}
\end{figure}

We explored the effect of electron doping by performing a gradual in-situ potassium (K) deposition. Doping created an interesting spectral weight redistribution, as shown in Figure \ref{FigureDop}. As we doped with electrons, weight grew significantly at the V 3d identified band, while increasing at lower amounts or staying the same at higher binding energies. 
The amount of K dosed on the surface of VI$_3$ had a very weak impact on the spectral weight at deep binding energies (below 7 eV). In contrast, the top of the valence band showed in addition to an important grow in weight, a noticeable broadening and a modest shift to higher binding energies as more K was deposited. This behavior is in sharp contrast with the effect of in-situ doping on semiconductors or even Mott insulators, where the K-donated electrons usually trigger a rigid shift of the valence bands toward higher binding energies  with no spectral weight redistribution. In those cases, an energy band gap can be estimated from the consequential shift of the chemical potential into the conduction band \cite{PhysRevLett.109.266406}.  Here, the observed change of the band structure with doping suggests a non-conventional evolution of the VI$_3$ Mott state with doping.  

Using different polarizations of the synchrotron light in ARPES is known to reveal hidden traits in the electronic band structure, or to help identifying the orbital character of materials' band manifolds \cite{doi:10.1021/acsnano.9b04536, PhysRevB.85.214518, PhysRevB.100.241406, Cao2013}. As observed in Figure \ref{Figure2}, in our setup (Beamline 4.0.3 at the ALS) the high energetic light is incident in the $k_y$=0 plane (65$^{\circ}$ relative to the analyzer lens), which breaks any geometrical symmetry about the crystal $k_x$=0 plane. As a consequence, there is no symmetry in the photoemission measured between $\pm k_x$ points for different light polarizations, and in contrast there is a generic geometrical dichroism (difference in the photoemission for two opposite light polarizations) above and below the $k_y=0$ incident plane (at $\pm k_y$ for any $k_x$), provided that the crystal has also a symmetry above and below the $k_y=0$ plane. In this study, we make use of linearly and circularly polarized light. In a first instance, we use out-of-plane $\perp$ (or LH) and in-plane $\parallel$ (or LV) light polarizations. In the case of $\perp$ light, the electric field vector of the incident photons has a perpendicular component with respect to the sample surface, while for of $\parallel$ light, the electric field is always parallel to the sample surface (see Figure \ref{Figure2}e).      
Figures \ref{Figure2}a and b show VI$_3$ ARPES measured for both in-plane $\parallel$ and out-of-plane $\perp$ light polarizations in the K-$\Gamma$-K direction, as indicated in Figure \ref{Figure1}c. We observe that different bands are enhanced with the two light polarizations. In particular, $\perp$ light polarization reveals a stronger contribution from the band near $1$ eV below the Fermi energy with respect to other bands at higher binding energies. This is summarized in Figure \ref{Figure2}c, where the difference in ARPES intensity using the two polarizations (blue $\perp$, red $\parallel$) known as linear dichroism, is calculated. 
We have represented in Figure \ref{Figure2}d the momentum-integrated energy distribution curves (EDCs) for the $\perp$ (blue) and $\parallel$ (red) polarizations, normalized by the incident photon flux. We observe that the intensity of the band near $1$ eV below the Fermi energy is appreciably reduced for the $\parallel$ with respect to the $\perp$ light polarization (by 21$\%$). 
Our linear dichroism data shows therefore that the V-3d-electron-derived bands are particularly sensitive to $\perp$ light polarization, indicating that their orbitals are ordered mainly orthogonal to the sample surface, while bands at higher binding energies (associated to I 5p orbitals \cite{Kundu2020}) seem to be ordered mainly parallel to the sample surface (ab plane). The former bands host the t$_{2g}$ orbitals that set the Mott insulating behavior in VI$_3$ as well as its magnetic properties. As mentioned in the introduction, two possible scenarios explain the electronic properties of VI$_3$. One in which V d$^{2}$ electrons occupy the $|e_\pm'\rangle$ orbitals and another where the V d$^{2}$ electrons occupy the $|a_{1g}\rangle$ and $|e_-'\rangle$ orbitals. The real part of the $|e_\pm'\rangle$ orbitals can be represented in coordinates with the z axis along the [111] axes of the crystal ($|e_{g1}'\rangle$ and $|e_{g2}'\rangle$), where they lie in the x-y plane (with some tilting of $|e_{g2}'\rangle$ with respect to $|e_{g1}'\rangle$) \cite{khomskii_2014}
\begin{align*}
|e'_{g1}\rangle&=\frac{1}{\sqrt{3}}\big(\sqrt{2}(x^2-y^2)-xz\big)\\
|e'_{g2}\rangle&=\frac{1}{\sqrt{3}}\big(\sqrt{2}xy+yz\big).
\end{align*}
On the other hand, the $|a_{1g}\rangle$ orbitals are real and oriented parallel to the z direction \cite{khomskii_2014, PhysRevB.104.014414},
$$
|a_{1g}\rangle\approx 3z^2-r^2.
$$
The measured linear dichroism for the V 3d bands points towards orbitals oriented mostly out-of-plane and therefore to a preferred $a_{1g}^1e_{\pm}^{1}$ ground state for VI$_3$, instead of a $e_{\pm}^{2}a_{1g}^0$ ground state where orbitals are expected to be fully on-plane.     

\begin{figure}[H]
\includegraphics[width=14cm]{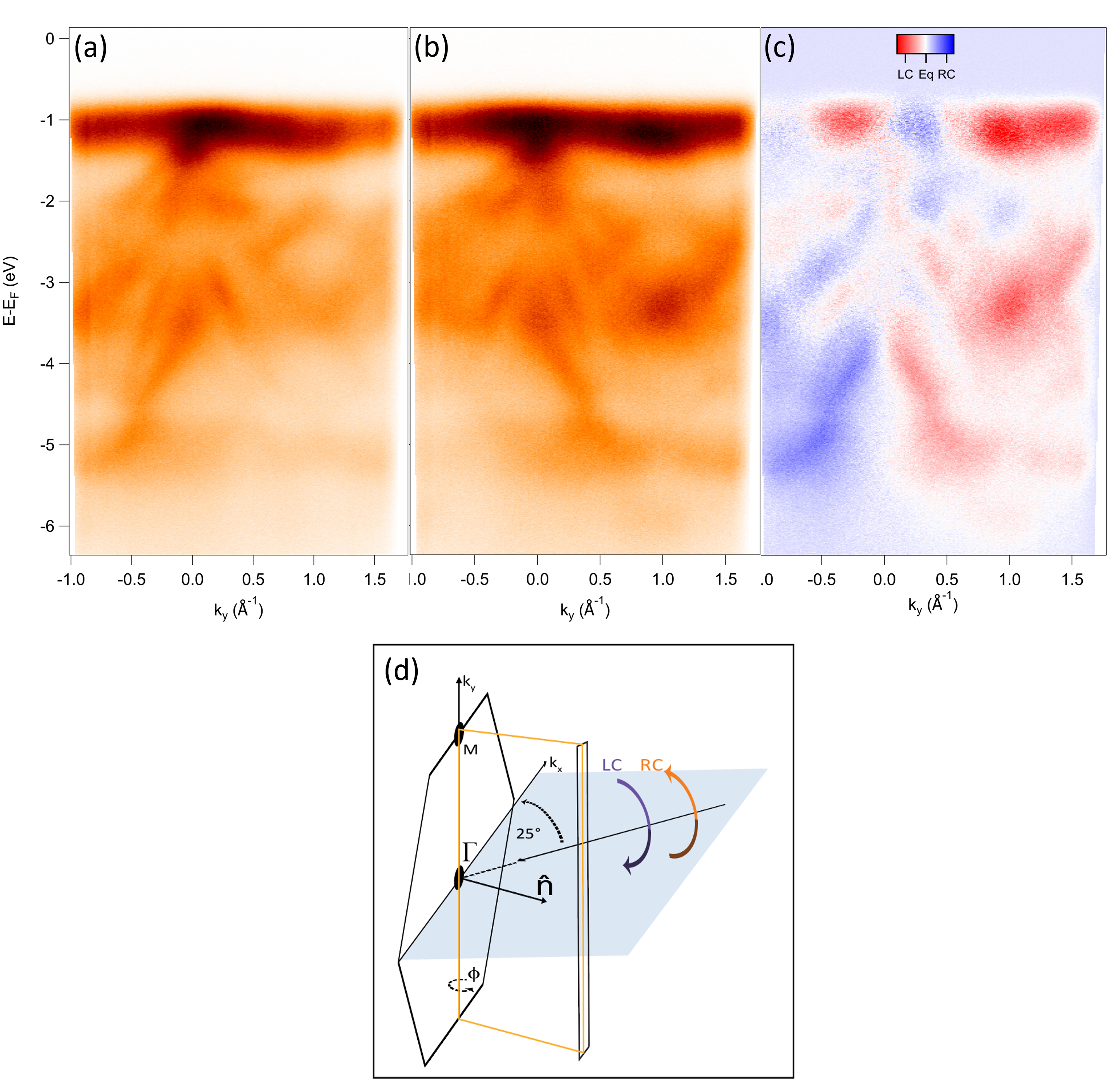}\caption{\label{Figure3}\textbf{Circular dichroism of VI$_3$ ARPES} a) VI$_3$ ARPES using left-hand circularly (LC) (a) and right-hand circularly polarized (RC) light (b) along the K-$\Gamma$-K direction. (c) Circular dichroism calculated using (a) and (b). (d) Schematics of the experimental setup as in figure \ref{Figure2}e showing incident photons right-hand (red) and left-hand (blue) circularly polarized. }
\end{figure}

In order to explore further the orbital character details of the VI$_3$ valence bands, we studied the ARPES circular dichroism in the photoelectron angular distribution (CDAD), by using  left- (LC) and right- hand (RC) circularly polarized light. In general, CDAD can be observed in an experimental geometry with a handedness. CDAD is absent when the geometry is such that a symmetry operation that transforms right- into left-circularly polarized light leaves the momentum vector of the photoelectrons \textbf{k} unaffected \cite{Rader}. This is the case of the $k_y=0$ plane (in blue in Figure \ref{Figure3}d), that transforms right into left circularly polarized light without affecting \textbf{k}, parallel to $\hat{n}$, the normal to the sample. There is therefore zero CDAD for $k_y$=0. On the other hand on the detector plane (yellow plane in Figure \ref{Figure3}d), a reversal of the light polarization changes the sign of the y component of \textbf{k}, leading to a non-zero CDAD. 
In our data, the CDAD is zero at $\Gamma$, as observed in Figure \ref{Figure3}c. The handedness of the experimental setup combined with the symmetry of the orbitals making up the band structure of VI$_3$, result in an CDAD intensity with opposite sign when crossing $\Gamma$ along $k_y$. Despite the fact that CDAD has been argued to be a final state effect \cite{Rader,Bigi}, it is worth it to point out that here, the V 3d identified bands exhibit the same CDAD polarization, while bands at higher binding energies (below 1 eV) show a switching of the CDAD. This is in contrast with the linear dichroism, that shows the same important LV light polarization effect at binding energies below 1 eV.

\begin{figure}[H]
\includegraphics[width=10cm]{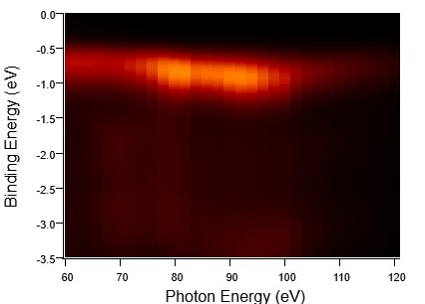}\caption{\label{Figure4}\textbf{Photon energy dependence of the photoemision} along the K-$\Gamma$-K direction in the photon energy range 60 - 120 eV.}
\end{figure}

Finally, we study the photon energy dependence of VI$_3$ ARPES. Figure \ref{Figure4} shows data taken along the K-$\Gamma$-K direction with $\perp$ photons in a photon energy range of 60-120 eV. The top of the VI$_3$ valence band stays at about the same energy position for different photon energies, indicating a very weak band dispersion along the k$_z$ direction, characteristic of layered materials with no interacting layers.

In summary, using different light polarizations, our ARPES study on the vdW Mott ferromagnet VI$_3$ points towards a ground state where V d$^2$ electrons occupy the $|a_{1g}\rangle$ and half fill the $|e_\pm'\rangle$ state.
Circularly polarized light shows the same dichroism (CDAD) for the identified V 3d states while alternating CDAD at higher binding energies, indicating the same orbital character of the V 3d bands near the top of the conduction band. Our photon energy dependent measurements confirm the layer character of VI$_3$, with weakly interacting layers. Most importantly, our measurements put in evidence in an indirect way the importance of spin-orbit coupling in this vdW ferromagnet, that leads to the $a_{1g}^1e_{\pm}^{1}$ state, expected to be metallic unless spin-orbit coupling and electronic correlations are active and open a Mott band gap. As detailed in the introduction, spin orbit coupling in VI$_3$ plays indeed an essential role in the observed perpendicular magnetic anisotropy, essential for electronic applications.

\begin{acknowledgments}
The primary funding of this work was provided by the U.S. Department of Energy, Office of Science, Office of Basic Energy Sciences under contract DE-SC0018154. The crystal growth work conducted at Princeton University was supported by the NSF-sponsored PARADIGM program centered at Cornell University, grant DMR-1539918 We would like to acknowledge invaluable advice from Jonathan Denlinger from the Advanced Light Source at LBL as well as important discussions with Peng Zhang and calculations from Elton J. G. Santos, Benjamin Carpenter and Ignacio Martin Alliati.
\end{acknowledgments}

\bibliography{Ojeda-AristizabalVI3}

\begin{thebibliography}{27}%
\makeatletter
\providecommand \@ifxundefined [1]{%
 \@ifx{#1\undefined}
}%
\providecommand \@ifnum [1]{%
 \ifnum #1\expandafter \@firstoftwo
 \else \expandafter \@secondoftwo
 \fi
}%
\providecommand \@ifx [1]{%
 \ifx #1\expandafter \@firstoftwo
 \else \expandafter \@secondoftwo
 \fi
}%
\providecommand \natexlab [1]{#1}%
\providecommand \enquote  [1]{``#1''}%
\providecommand \bibnamefont  [1]{#1}%
\providecommand \bibfnamefont [1]{#1}%
\providecommand \citenamefont [1]{#1}%
\providecommand \href@noop [0]{\@secondoftwo}%
\providecommand \href [0]{\begingroup \@sanitize@url \@href}%
\providecommand \@href[1]{\@@startlink{#1}\@@href}%
\providecommand \@@href[1]{\endgroup#1\@@endlink}%
\providecommand \@sanitize@url [0]{\catcode `\\12\catcode `\$12\catcode
  `\&12\catcode `\#12\catcode `\^12\catcode `\_12\catcode `\%12\relax}%
\providecommand \@@startlink[1]{}%
\providecommand \@@endlink[0]{}%
\providecommand \url  [0]{\begingroup\@sanitize@url \@url }%
\providecommand \@url [1]{\endgroup\@href {#1}{\urlprefix }}%
\providecommand \urlprefix  [0]{URL }%
\providecommand \Eprint [0]{\href }%
\providecommand \doibase [0]{http://dx.doi.org/}%
\providecommand \selectlanguage [0]{\@gobble}%
\providecommand \bibinfo  [0]{\@secondoftwo}%
\providecommand \bibfield  [0]{\@secondoftwo}%
\providecommand \translation [1]{[#1]}%
\providecommand \BibitemOpen [0]{}%
\providecommand \bibitemStop [0]{}%
\providecommand \bibitemNoStop [0]{.\EOS\space}%
\providecommand \EOS [0]{\spacefactor3000\relax}%
\providecommand \BibitemShut  [1]{\csname bibitem#1\endcsname}%
\let\auto@bib@innerbib\@empty
\bibitem [{\citenamefont {Lin}\ \emph {et~al.}(2021)\citenamefont {Lin},
  \citenamefont {Huang}, \citenamefont {Hwangbo}, \citenamefont {Jiang},
  \citenamefont {Zhang}, \citenamefont {Liu}, \citenamefont {Fei},
  \citenamefont {Lv}, \citenamefont {Millis}, \citenamefont {McGuire},
  \citenamefont {Xiao}, \citenamefont {Chu},\ and\ \citenamefont
  {Xu}}]{doi:10.1021/acs.nanolett.1c03027}%
  \BibitemOpen
  \bibfield  {author} {\bibinfo {author} {\bibfnamefont {Z.}~\bibnamefont
  {Lin}}, \bibinfo {author} {\bibfnamefont {B.}~\bibnamefont {Huang}}, \bibinfo
  {author} {\bibfnamefont {K.}~\bibnamefont {Hwangbo}}, \bibinfo {author}
  {\bibfnamefont {Q.}~\bibnamefont {Jiang}}, \bibinfo {author} {\bibfnamefont
  {Q.}~\bibnamefont {Zhang}}, \bibinfo {author} {\bibfnamefont
  {Z.}~\bibnamefont {Liu}}, \bibinfo {author} {\bibfnamefont {Z.}~\bibnamefont
  {Fei}}, \bibinfo {author} {\bibfnamefont {H.}~\bibnamefont {Lv}}, \bibinfo
  {author} {\bibfnamefont {A.}~\bibnamefont {Millis}}, \bibinfo {author}
  {\bibfnamefont {M.}~\bibnamefont {McGuire}}, \bibinfo {author} {\bibfnamefont
  {D.}~\bibnamefont {Xiao}}, \bibinfo {author} {\bibfnamefont {J.-H.}\
  \bibnamefont {Chu}}, \ and\ \bibinfo {author} {\bibfnamefont
  {X.}~\bibnamefont {Xu}},\ }\bibfield  {title} {\enquote {\bibinfo {title}
  {Magnetism and its structural coupling effects in 2d ising ferromagnetic
  insulator vi3},}\ }\href {\doibase 10.1021/acs.nanolett.1c03027} {\bibfield
  {journal} {\bibinfo  {journal} {Nano Letters}\ }\textbf {\bibinfo {volume}
  {21}},\ \bibinfo {pages} {9180--9186} (\bibinfo {year} {2021})},\ \bibinfo
  {note} {pMID: 34724786},\ \Eprint
  {http://arxiv.org/abs/https://doi.org/10.1021/acs.nanolett.1c03027}
  {https://doi.org/10.1021/acs.nanolett.1c03027} \BibitemShut {NoStop}%
\bibitem [{\citenamefont {Khomskii}(2014)}]{khomskii_2014}%
  \BibitemOpen
  \bibfield  {author} {\bibinfo {author} {\bibfnamefont {D.~I.}\ \bibnamefont
  {Khomskii}},\ }\href {\doibase 10.1017/CBO9781139096782} {\emph {\bibinfo
  {title} {Transition Metal Compounds}}}\ (\bibinfo  {publisher} {Cambridge
  University Press},\ \bibinfo {year} {2014})\BibitemShut {NoStop}%
\bibitem [{\citenamefont {Nguyen}\ \emph {et~al.}(2021)\citenamefont {Nguyen},
  \citenamefont {Yamauchi}, \citenamefont {Oguchi}, \citenamefont {Amoroso},\
  and\ \citenamefont {Picozzi}}]{PhysRevB.104.014414}%
  \BibitemOpen
  \bibfield  {author} {\bibinfo {author} {\bibfnamefont {T.~P.~T.}\
  \bibnamefont {Nguyen}}, \bibinfo {author} {\bibfnamefont {K.}~\bibnamefont
  {Yamauchi}}, \bibinfo {author} {\bibfnamefont {T.}~\bibnamefont {Oguchi}},
  \bibinfo {author} {\bibfnamefont {D.}~\bibnamefont {Amoroso}}, \ and\
  \bibinfo {author} {\bibfnamefont {S.}~\bibnamefont {Picozzi}},\ }\bibfield
  {title} {\enquote {\bibinfo {title} {Electric-field tuning of the magnetic
  properties of bilayer ${\mathrm{vi}}_{3}$: A first-principles study},}\
  }\href {\doibase 10.1103/PhysRevB.104.014414} {\bibfield  {journal} {\bibinfo
   {journal} {Phys. Rev. B}\ }\textbf {\bibinfo {volume} {104}},\ \bibinfo
  {pages} {014414} (\bibinfo {year} {2021})}\BibitemShut {NoStop}%
\bibitem [{\citenamefont {Wang}\ and\ \citenamefont
  {Long}(2020)}]{PhysRevB.101.024411}%
  \BibitemOpen
  \bibfield  {author} {\bibinfo {author} {\bibfnamefont {Y.-P.}\ \bibnamefont
  {Wang}}\ and\ \bibinfo {author} {\bibfnamefont {M.-Q.}\ \bibnamefont
  {Long}},\ }\bibfield  {title} {\enquote {\bibinfo {title} {Electronic and
  magnetic properties of van der waals ferromagnetic semiconductor
  ${\mathrm{vi}}_{3}$},}\ }\href {\doibase 10.1103/PhysRevB.101.024411}
  {\bibfield  {journal} {\bibinfo  {journal} {Phys. Rev. B}\ }\textbf {\bibinfo
  {volume} {101}},\ \bibinfo {pages} {024411} (\bibinfo {year}
  {2020})}\BibitemShut {NoStop}%
\bibitem [{\citenamefont {Huang}\ \emph {et~al.}(2020)\citenamefont {Huang},
  \citenamefont {Wu}, \citenamefont {Yu}, \citenamefont {Jena},\ and\
  \citenamefont {Kan}}]{C9CP05643B}%
  \BibitemOpen
  \bibfield  {author} {\bibinfo {author} {\bibfnamefont {C.}~\bibnamefont
  {Huang}}, \bibinfo {author} {\bibfnamefont {F.}~\bibnamefont {Wu}}, \bibinfo
  {author} {\bibfnamefont {S.}~\bibnamefont {Yu}}, \bibinfo {author}
  {\bibfnamefont {P.}~\bibnamefont {Jena}}, \ and\ \bibinfo {author}
  {\bibfnamefont {E.}~\bibnamefont {Kan}},\ }\bibfield  {title} {\enquote
  {\bibinfo {title} {Discovery of twin orbital-order phases in ferromagnetic
  semiconducting vi3 monolayer},}\ }\href {\doibase 10.1039/C9CP05643B}
  {\bibfield  {journal} {\bibinfo  {journal} {Phys. Chem. Chem. Phys.}\
  }\textbf {\bibinfo {volume} {22}},\ \bibinfo {pages} {512--517} (\bibinfo
  {year} {2020})}\BibitemShut {NoStop}%
\bibitem [{\citenamefont {Yang}\ \emph {et~al.}(2020)\citenamefont {Yang},
  \citenamefont {Fan}, \citenamefont {Wang}, \citenamefont {Khomskii},\ and\
  \citenamefont {Wu}}]{PhysRevB.101.100402}%
  \BibitemOpen
  \bibfield  {author} {\bibinfo {author} {\bibfnamefont {K.}~\bibnamefont
  {Yang}}, \bibinfo {author} {\bibfnamefont {F.}~\bibnamefont {Fan}}, \bibinfo
  {author} {\bibfnamefont {H.}~\bibnamefont {Wang}}, \bibinfo {author}
  {\bibfnamefont {D.~I.}\ \bibnamefont {Khomskii}}, \ and\ \bibinfo {author}
  {\bibfnamefont {H.}~\bibnamefont {Wu}},\ }\bibfield  {title} {\enquote
  {\bibinfo {title} {${\mathrm{vi}}_{3}$: A two-dimensional ising
  ferromagnet},}\ }\href {\doibase 10.1103/PhysRevB.101.100402} {\bibfield
  {journal} {\bibinfo  {journal} {Phys. Rev. B}\ }\textbf {\bibinfo {volume}
  {101}},\ \bibinfo {pages} {100402} (\bibinfo {year} {2020})}\BibitemShut
  {NoStop}%
\bibitem [{\citenamefont {Mermin}\ and\ \citenamefont
  {Wagner}(1966)}]{MerminWagnerFerro}%
  \BibitemOpen
  \bibfield  {author} {\bibinfo {author} {\bibfnamefont {N.~D.}\ \bibnamefont
  {Mermin}}\ and\ \bibinfo {author} {\bibfnamefont {H.}~\bibnamefont
  {Wagner}},\ }\bibfield  {title} {\enquote {\bibinfo {title} {Absence of
  ferromagnetism or antiferromagnetism in one- or two-dimensional isotropic
  heisenberg models},}\ }\href {\doibase 10.1103/PhysRevLett.17.1133}
  {\bibfield  {journal} {\bibinfo  {journal} {Phys. Rev. Lett.}\ }\textbf
  {\bibinfo {volume} {17}},\ \bibinfo {pages} {1133--1136} (\bibinfo {year}
  {1966})}\BibitemShut {NoStop}%
\bibitem [{\citenamefont {Huang}\ \emph {et~al.}(2017)\citenamefont {Huang},
  \citenamefont {Clark}, \citenamefont {Navarro-Moratalla}, \citenamefont
  {Klein}, \citenamefont {Cheng}, \citenamefont {Seyler}, \citenamefont
  {Zhong}, \citenamefont {Schmidgall}, \citenamefont {McGuire}, \citenamefont
  {Cobden}, \citenamefont {Yao}, \citenamefont {Xiao}, \citenamefont
  {Jarillo-Herrero},\ and\ \citenamefont {Xu}}]{Huang2017}%
  \BibitemOpen
  \bibfield  {author} {\bibinfo {author} {\bibfnamefont {B.}~\bibnamefont
  {Huang}}, \bibinfo {author} {\bibfnamefont {G.}~\bibnamefont {Clark}},
  \bibinfo {author} {\bibfnamefont {E.}~\bibnamefont {Navarro-Moratalla}},
  \bibinfo {author} {\bibfnamefont {D.~R.}\ \bibnamefont {Klein}}, \bibinfo
  {author} {\bibfnamefont {R.}~\bibnamefont {Cheng}}, \bibinfo {author}
  {\bibfnamefont {K.~L.}\ \bibnamefont {Seyler}}, \bibinfo {author}
  {\bibfnamefont {D.}~\bibnamefont {Zhong}}, \bibinfo {author} {\bibfnamefont
  {E.}~\bibnamefont {Schmidgall}}, \bibinfo {author} {\bibfnamefont {M.~A.}\
  \bibnamefont {McGuire}}, \bibinfo {author} {\bibfnamefont {D.~H.}\
  \bibnamefont {Cobden}}, \bibinfo {author} {\bibfnamefont {W.}~\bibnamefont
  {Yao}}, \bibinfo {author} {\bibfnamefont {D.}~\bibnamefont {Xiao}}, \bibinfo
  {author} {\bibfnamefont {P.}~\bibnamefont {Jarillo-Herrero}}, \ and\ \bibinfo
  {author} {\bibfnamefont {X.}~\bibnamefont {Xu}},\ }\bibfield  {title}
  {\enquote {\bibinfo {title} {Layer-dependent ferromagnetism in a van der
  waals crystal down to the monolayer limit},}\ }\href {\doibase
  10.1038/nature22391} {\bibfield  {journal} {\bibinfo  {journal} {Nature}\
  }\textbf {\bibinfo {volume} {546}},\ \bibinfo {pages} {270--273} (\bibinfo
  {year} {2017})}\BibitemShut {NoStop}%
\bibitem [{\citenamefont {Gong}\ \emph {et~al.}(2017)\citenamefont {Gong},
  \citenamefont {Li}, \citenamefont {Li}, \citenamefont {Ji}, \citenamefont
  {Stern}, \citenamefont {Xia}, \citenamefont {Cao}, \citenamefont {Bao},
  \citenamefont {Wang}, \citenamefont {Wang}, \citenamefont {Qiu},
  \citenamefont {Cava}, \citenamefont {Louie}, \citenamefont {Xia},\ and\
  \citenamefont {Zhang}}]{Gong2017}%
  \BibitemOpen
  \bibfield  {author} {\bibinfo {author} {\bibfnamefont {C.}~\bibnamefont
  {Gong}}, \bibinfo {author} {\bibfnamefont {L.}~\bibnamefont {Li}}, \bibinfo
  {author} {\bibfnamefont {Z.}~\bibnamefont {Li}}, \bibinfo {author}
  {\bibfnamefont {H.}~\bibnamefont {Ji}}, \bibinfo {author} {\bibfnamefont
  {A.}~\bibnamefont {Stern}}, \bibinfo {author} {\bibfnamefont
  {Y.}~\bibnamefont {Xia}}, \bibinfo {author} {\bibfnamefont {T.}~\bibnamefont
  {Cao}}, \bibinfo {author} {\bibfnamefont {W.}~\bibnamefont {Bao}}, \bibinfo
  {author} {\bibfnamefont {C.}~\bibnamefont {Wang}}, \bibinfo {author}
  {\bibfnamefont {Y.}~\bibnamefont {Wang}}, \bibinfo {author} {\bibfnamefont
  {Z.~Q.}\ \bibnamefont {Qiu}}, \bibinfo {author} {\bibfnamefont {R.~J.}\
  \bibnamefont {Cava}}, \bibinfo {author} {\bibfnamefont {S.~G.}\ \bibnamefont
  {Louie}}, \bibinfo {author} {\bibfnamefont {J.}~\bibnamefont {Xia}}, \ and\
  \bibinfo {author} {\bibfnamefont {X.}~\bibnamefont {Zhang}},\ }\bibfield
  {title} {\enquote {\bibinfo {title} {Discovery of intrinsic ferromagnetism in
  two-dimensional van der waals crystals},}\ }\href {\doibase
  10.1038/nature22060} {\bibfield  {journal} {\bibinfo  {journal} {Nature}\
  }\textbf {\bibinfo {volume} {546}},\ \bibinfo {pages} {265--296} (\bibinfo
  {year} {2017})}\BibitemShut {NoStop}%
\bibitem [{\citenamefont {Lado}\ and\ \citenamefont
  {Fern{\'{a}}ndez-Rossier}(2017)}]{Lado_2017}%
  \BibitemOpen
  \bibfield  {author} {\bibinfo {author} {\bibfnamefont {J.~L.}\ \bibnamefont
  {Lado}}\ and\ \bibinfo {author} {\bibfnamefont {J.}~\bibnamefont
  {Fern{\'{a}}ndez-Rossier}},\ }\bibfield  {title} {\enquote {\bibinfo {title}
  {On the origin of magnetic anisotropy in two dimensional cri$_3$},}\ }\href
  {\doibase 10.1088/2053-1583/aa75ed} {\bibfield  {journal} {\bibinfo
  {journal} {2D Materials}\ }\textbf {\bibinfo {volume} {4}},\ \bibinfo {pages}
  {035002} (\bibinfo {year} {2017})}\BibitemShut {NoStop}%
\bibitem [{\citenamefont {Kim}\ \emph {et~al.}(2019)\citenamefont {Kim},
  \citenamefont {Kim}, \citenamefont {Ko}, \citenamefont {Seo}, \citenamefont
  {Kim}, \citenamefont {Jang}, \citenamefont {Kim}, \citenamefont {Kim},
  \citenamefont {Cheong},\ and\ \citenamefont {Park}}]{PhysRevLett.122.207201}%
  \BibitemOpen
  \bibfield  {author} {\bibinfo {author} {\bibfnamefont {D.-H.}\ \bibnamefont
  {Kim}}, \bibinfo {author} {\bibfnamefont {K.}~\bibnamefont {Kim}}, \bibinfo
  {author} {\bibfnamefont {K.-T.}\ \bibnamefont {Ko}}, \bibinfo {author}
  {\bibfnamefont {J.}~\bibnamefont {Seo}}, \bibinfo {author} {\bibfnamefont
  {J.~S.}\ \bibnamefont {Kim}}, \bibinfo {author} {\bibfnamefont {T.-H.}\
  \bibnamefont {Jang}}, \bibinfo {author} {\bibfnamefont {Y.}~\bibnamefont
  {Kim}}, \bibinfo {author} {\bibfnamefont {J.-Y.}\ \bibnamefont {Kim}},
  \bibinfo {author} {\bibfnamefont {S.-W.}\ \bibnamefont {Cheong}}, \ and\
  \bibinfo {author} {\bibfnamefont {J.-H.}\ \bibnamefont {Park}},\ }\bibfield
  {title} {\enquote {\bibinfo {title} {Giant magnetic anisotropy induced by
  ligand $ls$ coupling in layered cr compounds},}\ }\href {\doibase
  10.1103/PhysRevLett.122.207201} {\bibfield  {journal} {\bibinfo  {journal}
  {Phys. Rev. Lett.}\ }\textbf {\bibinfo {volume} {122}},\ \bibinfo {pages}
  {207201} (\bibinfo {year} {2019})}\BibitemShut {NoStop}%
\bibitem [{\citenamefont {Koriki}\ \emph {et~al.}(2021)\citenamefont {Koriki},
  \citenamefont {M\'{\i}\ifmmode~\check{s}\else \v{s}\fi{}ek}, \citenamefont
  {Posp\'{\i}\ifmmode~\check{s}\else \v{s}\fi{}il}, \citenamefont
  {Kratochv\'{\i}lov\'a}, \citenamefont {Carva}, \citenamefont
  {Prokle\ifmmode~\check{s}\else \v{s}\fi{}ka}, \citenamefont
  {Dole\ifmmode~\check{z}\else \v{z}\fi{}al}, \citenamefont
  {Ka\ifmmode~\check{s}\else \v{s}\fi{}til}, \citenamefont {Son}, \citenamefont
  {Park},\ and\ \citenamefont {Sechovsk\'y}}]{PhysRevB.103.174401}%
  \BibitemOpen
  \bibfield  {author} {\bibinfo {author} {\bibfnamefont {A.}~\bibnamefont
  {Koriki}}, \bibinfo {author} {\bibfnamefont {M.}~\bibnamefont
  {M\'{\i}\ifmmode~\check{s}\else \v{s}\fi{}ek}}, \bibinfo {author}
  {\bibfnamefont {J.}~\bibnamefont {Posp\'{\i}\ifmmode~\check{s}\else
  \v{s}\fi{}il}}, \bibinfo {author} {\bibfnamefont {M.}~\bibnamefont
  {Kratochv\'{\i}lov\'a}}, \bibinfo {author} {\bibfnamefont {K.}~\bibnamefont
  {Carva}}, \bibinfo {author} {\bibfnamefont {J.}~\bibnamefont
  {Prokle\ifmmode~\check{s}\else \v{s}\fi{}ka}}, \bibinfo {author}
  {\bibfnamefont {P.}~\bibnamefont {Dole\ifmmode~\check{z}\else \v{z}\fi{}al}},
  \bibinfo {author} {\bibfnamefont {J.}~\bibnamefont {Ka\ifmmode~\check{s}\else
  \v{s}\fi{}til}}, \bibinfo {author} {\bibfnamefont {S.}~\bibnamefont {Son}},
  \bibinfo {author} {\bibfnamefont {J.-G.}\ \bibnamefont {Park}}, \ and\
  \bibinfo {author} {\bibfnamefont {V.}~\bibnamefont {Sechovsk\'y}},\
  }\bibfield  {title} {\enquote {\bibinfo {title} {Magnetic anisotropy in the
  van der waals ferromagnet $\mathrm{V}{\mathrm{i}}_{3}$},}\ }\href {\doibase
  10.1103/PhysRevB.103.174401} {\bibfield  {journal} {\bibinfo  {journal}
  {Phys. Rev. B}\ }\textbf {\bibinfo {volume} {103}},\ \bibinfo {pages}
  {174401} (\bibinfo {year} {2021})}\BibitemShut {NoStop}%
\bibitem [{\citenamefont {Kundu}\ \emph {et~al.}(2020)\citenamefont {Kundu},
  \citenamefont {Liu}, \citenamefont {Petrovic},\ and\ \citenamefont
  {Valla}}]{Kundu2020}%
  \BibitemOpen
  \bibfield  {author} {\bibinfo {author} {\bibfnamefont {A.~K.}\ \bibnamefont
  {Kundu}}, \bibinfo {author} {\bibfnamefont {Y.}~\bibnamefont {Liu}}, \bibinfo
  {author} {\bibfnamefont {C.}~\bibnamefont {Petrovic}}, \ and\ \bibinfo
  {author} {\bibfnamefont {T.}~\bibnamefont {Valla}},\ }\bibfield  {title}
  {\enquote {\bibinfo {title} {Valence band electronic structure of the van der
  waals ferromagnetic insulators: Vi3and cri3},}\ }\href {\doibase
  10.1038/s41598-020-72487-5} {\bibfield  {journal} {\bibinfo  {journal}
  {Scientific Report}\ }\textbf {\bibinfo {volume} {10}} (\bibinfo {year}
  {2020}),\ 10.1038/s41598-020-72487-5}\BibitemShut {NoStop}%
\bibitem [{\citenamefont {Kong}\ \emph {et~al.}(2019)\citenamefont {Kong},
  \citenamefont {Stolze}, \citenamefont {Timmons}, \citenamefont {Tao},
  \citenamefont {Ni}, \citenamefont {Guo}, \citenamefont {Yang}, \citenamefont
  {Prozorov},\ and\ \citenamefont {Cava}}]{TaiKong}%
  \BibitemOpen
  \bibfield  {author} {\bibinfo {author} {\bibfnamefont {T.}~\bibnamefont
  {Kong}}, \bibinfo {author} {\bibfnamefont {K.}~\bibnamefont {Stolze}},
  \bibinfo {author} {\bibfnamefont {E.~I.}\ \bibnamefont {Timmons}}, \bibinfo
  {author} {\bibfnamefont {J.}~\bibnamefont {Tao}}, \bibinfo {author}
  {\bibfnamefont {D.}~\bibnamefont {Ni}}, \bibinfo {author} {\bibfnamefont
  {S.}~\bibnamefont {Guo}}, \bibinfo {author} {\bibfnamefont {Z.}~\bibnamefont
  {Yang}}, \bibinfo {author} {\bibfnamefont {R.}~\bibnamefont {Prozorov}}, \
  and\ \bibinfo {author} {\bibfnamefont {R.~J.}\ \bibnamefont {Cava}},\
  }\bibfield  {title} {\enquote {\bibinfo {title} {Vi3—a new layered
  ferromagnetic semiconductor},}\ }\href {\doibase
  https://doi.org/10.1002/adma.201808074} {\bibfield  {journal} {\bibinfo
  {journal} {Advanced Materials}\ }\textbf {\bibinfo {volume} {31}},\ \bibinfo
  {pages} {1808074} (\bibinfo {year} {2019})},\ \Eprint
  {http://arxiv.org/abs/https://onlinelibrary.wiley.com/doi/pdf/10.1002/adma.201808074}
  {https://onlinelibrary.wiley.com/doi/pdf/10.1002/adma.201808074} \BibitemShut
  {NoStop}%
\bibitem [{\citenamefont {Zhang}\ \emph {et~al.}(2011)\citenamefont {Zhang},
  \citenamefont {Richard}, \citenamefont {Qian}, \citenamefont {Xu},
  \citenamefont {Dai},\ and\ \citenamefont {Ding}}]{doi:10.1063/1.3585113}%
  \BibitemOpen
  \bibfield  {author} {\bibinfo {author} {\bibfnamefont {P.}~\bibnamefont
  {Zhang}}, \bibinfo {author} {\bibfnamefont {P.}~\bibnamefont {Richard}},
  \bibinfo {author} {\bibfnamefont {T.}~\bibnamefont {Qian}}, \bibinfo {author}
  {\bibfnamefont {Y.-M.}\ \bibnamefont {Xu}}, \bibinfo {author} {\bibfnamefont
  {X.}~\bibnamefont {Dai}}, \ and\ \bibinfo {author} {\bibfnamefont
  {H.}~\bibnamefont {Ding}},\ }\bibfield  {title} {\enquote {\bibinfo {title}
  {A precise method for visualizing dispersive features in image plots},}\
  }\href {\doibase 10.1063/1.3585113} {\bibfield  {journal} {\bibinfo
  {journal} {Review of Scientific Instruments}\ }\textbf {\bibinfo {volume}
  {82}},\ \bibinfo {pages} {043712} (\bibinfo {year} {2011})},\ \Eprint
  {http://arxiv.org/abs/https://doi.org/10.1063/1.3585113}
  {https://doi.org/10.1063/1.3585113} \BibitemShut {NoStop}%
\bibitem [{\citenamefont {Dole\ifmmode~\check{z}\else \v{z}\fi{}al}\ \emph
  {et~al.}(2019)\citenamefont {Dole\ifmmode~\check{z}\else \v{z}\fi{}al},
  \citenamefont {Kratochv\'{\i}lov\'a}, \citenamefont {Hol\'y}, \citenamefont
  {\ifmmode~\check{C}\else \v{C}\fi{}erm\'ak}, \citenamefont {Sechovsk\'y},
  \citenamefont {Du\ifmmode~\check{s}\else \v{s}\fi{}ek}, \citenamefont
  {M\'{\i}\ifmmode~\check{s}\else \v{s}\fi{}ek}, \citenamefont {Chakraborty},
  \citenamefont {Noda}, \citenamefont {Son},\ and\ \citenamefont
  {Park}}]{PhysRevMaterials.3.121401}%
  \BibitemOpen
  \bibfield  {author} {\bibinfo {author} {\bibfnamefont {P.}~\bibnamefont
  {Dole\ifmmode~\check{z}\else \v{z}\fi{}al}}, \bibinfo {author} {\bibfnamefont
  {M.}~\bibnamefont {Kratochv\'{\i}lov\'a}}, \bibinfo {author} {\bibfnamefont
  {V.}~\bibnamefont {Hol\'y}}, \bibinfo {author} {\bibfnamefont
  {P.}~\bibnamefont {\ifmmode~\check{C}\else \v{C}\fi{}erm\'ak}}, \bibinfo
  {author} {\bibfnamefont {V.}~\bibnamefont {Sechovsk\'y}}, \bibinfo {author}
  {\bibfnamefont {M.}~\bibnamefont {Du\ifmmode~\check{s}\else \v{s}\fi{}ek}},
  \bibinfo {author} {\bibfnamefont {M.}~\bibnamefont
  {M\'{\i}\ifmmode~\check{s}\else \v{s}\fi{}ek}}, \bibinfo {author}
  {\bibfnamefont {T.}~\bibnamefont {Chakraborty}}, \bibinfo {author}
  {\bibfnamefont {Y.}~\bibnamefont {Noda}}, \bibinfo {author} {\bibfnamefont
  {S.}~\bibnamefont {Son}}, \ and\ \bibinfo {author} {\bibfnamefont {J.-G.}\
  \bibnamefont {Park}},\ }\bibfield  {title} {\enquote {\bibinfo {title}
  {Crystal structures and phase transitions of the van der waals ferromagnet
  $\mathrm{V}{\mathrm{i}}_{3}$},}\ }\href {\doibase
  10.1103/PhysRevMaterials.3.121401} {\bibfield  {journal} {\bibinfo  {journal}
  {Phys. Rev. Materials}\ }\textbf {\bibinfo {volume} {3}},\ \bibinfo {pages}
  {121401} (\bibinfo {year} {2019})}\BibitemShut {NoStop}%
\bibitem [{\citenamefont {Feng}\ \emph {et~al.}(2018)\citenamefont {Feng},
  \citenamefont {Biswas}, \citenamefont {Rajan}, \citenamefont {Watson},
  \citenamefont {Mazzola}, \citenamefont {Clark}, \citenamefont {Underwood},
  \citenamefont {Marković}, \citenamefont {McLaren}, \citenamefont {Hunter},
  \citenamefont {Burn}, \citenamefont {Duffy}, \citenamefont {Barua},
  \citenamefont {Balakrishnan}, \citenamefont {Bertran}, \citenamefont
  {Le~Fèvre}, \citenamefont {Kim}, \citenamefont {van~der Laan}, \citenamefont
  {Hesjedal}, \citenamefont {Wahl},\ and\ \citenamefont {King}}]{VSe2_ACSNano}%
  \BibitemOpen
  \bibfield  {author} {\bibinfo {author} {\bibfnamefont {J.}~\bibnamefont
  {Feng}}, \bibinfo {author} {\bibfnamefont {D.}~\bibnamefont {Biswas}},
  \bibinfo {author} {\bibfnamefont {A.}~\bibnamefont {Rajan}}, \bibinfo
  {author} {\bibfnamefont {M.~D.}\ \bibnamefont {Watson}}, \bibinfo {author}
  {\bibfnamefont {F.}~\bibnamefont {Mazzola}}, \bibinfo {author} {\bibfnamefont
  {O.~J.}\ \bibnamefont {Clark}}, \bibinfo {author} {\bibfnamefont
  {K.}~\bibnamefont {Underwood}}, \bibinfo {author} {\bibfnamefont
  {I.}~\bibnamefont {Marković}}, \bibinfo {author} {\bibfnamefont
  {M.}~\bibnamefont {McLaren}}, \bibinfo {author} {\bibfnamefont
  {A.}~\bibnamefont {Hunter}}, \bibinfo {author} {\bibfnamefont {D.~M.}\
  \bibnamefont {Burn}}, \bibinfo {author} {\bibfnamefont {L.~B.}\ \bibnamefont
  {Duffy}}, \bibinfo {author} {\bibfnamefont {S.}~\bibnamefont {Barua}},
  \bibinfo {author} {\bibfnamefont {G.}~\bibnamefont {Balakrishnan}}, \bibinfo
  {author} {\bibfnamefont {F.}~\bibnamefont {Bertran}}, \bibinfo {author}
  {\bibfnamefont {P.}~\bibnamefont {Le~Fèvre}}, \bibinfo {author}
  {\bibfnamefont {T.~K.}\ \bibnamefont {Kim}}, \bibinfo {author} {\bibfnamefont
  {G.}~\bibnamefont {van~der Laan}}, \bibinfo {author} {\bibfnamefont
  {T.}~\bibnamefont {Hesjedal}}, \bibinfo {author} {\bibfnamefont
  {P.}~\bibnamefont {Wahl}}, \ and\ \bibinfo {author} {\bibfnamefont
  {P.~D.~C.}\ \bibnamefont {King}},\ }\bibfield  {title} {\enquote {\bibinfo
  {title} {Electronic structure and enhanced charge-density wave order of
  monolayer vse2},}\ }\href {\doibase 10.1021/acs.nanolett.8b01649} {\bibfield
  {journal} {\bibinfo  {journal} {Nano Letters}\ }\textbf {\bibinfo {volume}
  {18}},\ \bibinfo {pages} {4493--4499} (\bibinfo {year} {2018})},\ \bibinfo
  {note} {pMID: 29912565},\ \Eprint
  {http://arxiv.org/abs/https://doi.org/10.1021/acs.nanolett.8b01649}
  {https://doi.org/10.1021/acs.nanolett.8b01649} \BibitemShut {NoStop}%
\bibitem [{\citenamefont {Wang}\ \emph {et~al.}(2019)\citenamefont {Wang},
  \citenamefont {Ren}, \citenamefont {Li}, \citenamefont {Wang}, \citenamefont
  {Peng}, \citenamefont {Yu}, \citenamefont {Duan},\ and\ \citenamefont
  {Zhou}}]{VTe2_PRB}%
  \BibitemOpen
  \bibfield  {author} {\bibinfo {author} {\bibfnamefont {Y.}~\bibnamefont
  {Wang}}, \bibinfo {author} {\bibfnamefont {J.}~\bibnamefont {Ren}}, \bibinfo
  {author} {\bibfnamefont {J.}~\bibnamefont {Li}}, \bibinfo {author}
  {\bibfnamefont {Y.}~\bibnamefont {Wang}}, \bibinfo {author} {\bibfnamefont
  {H.}~\bibnamefont {Peng}}, \bibinfo {author} {\bibfnamefont {P.}~\bibnamefont
  {Yu}}, \bibinfo {author} {\bibfnamefont {W.}~\bibnamefont {Duan}}, \ and\
  \bibinfo {author} {\bibfnamefont {S.}~\bibnamefont {Zhou}},\ }\bibfield
  {title} {\enquote {\bibinfo {title} {Evidence of charge density wave with
  anisotropic gap in a monolayer ${\mathrm{vte}}_{2}$ film},}\ }\href {\doibase
  10.1103/PhysRevB.100.241404} {\bibfield  {journal} {\bibinfo  {journal}
  {Phys. Rev. B}\ }\textbf {\bibinfo {volume} {100}},\ \bibinfo {pages}
  {241404} (\bibinfo {year} {2019})}\BibitemShut {NoStop}%
\bibitem [{\citenamefont {Parry}(1994)}]{CrossSectionParry}%
  \BibitemOpen
  \bibfield  {author} {\bibinfo {author} {\bibfnamefont {D.~E.}\ \bibnamefont
  {Parry}},\ }\bibfield  {title} {\enquote {\bibinfo {title} {Atomic
  calculation of photoionization cross-sections and asymmetry parameters j.-j.
  yeh, published by gordon and breach, langhorne pa, 1993 isbn 2-88124-585-4
  price \$125 (*\$65), £82 (*£42), ecu 104 (*ecu 54) (*special price for
  individuals who order direct from the publisher)},}\ }\href {\doibase
  https://doi.org/10.1002/rcm.1290080716} {\bibfield  {journal} {\bibinfo
  {journal} {Rapid Communications in Mass Spectrometry}\ }\textbf {\bibinfo
  {volume} {8}},\ \bibinfo {pages} {579--579} (\bibinfo {year} {1994})},\
  \Eprint
  {http://arxiv.org/abs/https://analyticalsciencejournals.onlinelibrary.wiley.com/doi/pdf/10.1002/rcm.1290080716}
  {https://analyticalsciencejournals.onlinelibrary.wiley.com/doi/pdf/10.1002/rcm.1290080716}
  \BibitemShut {NoStop}%
\bibitem [{\citenamefont {Yeh}\ and\ \citenamefont {Lindau}(1985)}]{YEH19851}%
  \BibitemOpen
  \bibfield  {author} {\bibinfo {author} {\bibfnamefont {J.}~\bibnamefont
  {Yeh}}\ and\ \bibinfo {author} {\bibfnamefont {I.}~\bibnamefont {Lindau}},\
  }\bibfield  {title} {\enquote {\bibinfo {title} {Atomic subshell
  photoionization cross sections and asymmetry parameters: 1 $leq$ z $leq$
  103},}\ }\href {\doibase https://doi.org/10.1016/0092-640X(85)90016-6}
  {\bibfield  {journal} {\bibinfo  {journal} {Atomic Data and Nuclear Data
  Tables}\ }\textbf {\bibinfo {volume} {32}},\ \bibinfo {pages} {1--155}
  (\bibinfo {year} {1985})}\BibitemShut {NoStop}%
\bibitem [{\citenamefont {Comin}\ \emph {et~al.}(2012)\citenamefont {Comin},
  \citenamefont {Levy}, \citenamefont {Ludbrook}, \citenamefont {Zhu},
  \citenamefont {Veenstra}, \citenamefont {Rosen}, \citenamefont {Singh},
  \citenamefont {Gegenwart}, \citenamefont {Stricker}, \citenamefont {Hancock},
  \citenamefont {van~der Marel}, \citenamefont {Elfimov},\ and\ \citenamefont
  {Damascelli}}]{PhysRevLett.109.266406}%
  \BibitemOpen
  \bibfield  {author} {\bibinfo {author} {\bibfnamefont {R.}~\bibnamefont
  {Comin}}, \bibinfo {author} {\bibfnamefont {G.}~\bibnamefont {Levy}},
  \bibinfo {author} {\bibfnamefont {B.}~\bibnamefont {Ludbrook}}, \bibinfo
  {author} {\bibfnamefont {Z.-H.}\ \bibnamefont {Zhu}}, \bibinfo {author}
  {\bibfnamefont {C.~N.}\ \bibnamefont {Veenstra}}, \bibinfo {author}
  {\bibfnamefont {J.~A.}\ \bibnamefont {Rosen}}, \bibinfo {author}
  {\bibfnamefont {Y.}~\bibnamefont {Singh}}, \bibinfo {author} {\bibfnamefont
  {P.}~\bibnamefont {Gegenwart}}, \bibinfo {author} {\bibfnamefont
  {D.}~\bibnamefont {Stricker}}, \bibinfo {author} {\bibfnamefont {J.~N.}\
  \bibnamefont {Hancock}}, \bibinfo {author} {\bibfnamefont {D.}~\bibnamefont
  {van~der Marel}}, \bibinfo {author} {\bibfnamefont {I.~S.}\ \bibnamefont
  {Elfimov}}, \ and\ \bibinfo {author} {\bibfnamefont {A.}~\bibnamefont
  {Damascelli}},\ }\bibfield  {title} {\enquote {\bibinfo {title}
  {${\mathrm{na}}_{2}{\mathrm{iro}}_{3}$ as a novel relativistic mott insulator
  with a 340-mev gap},}\ }\href {\doibase 10.1103/PhysRevLett.109.266406}
  {\bibfield  {journal} {\bibinfo  {journal} {Phys. Rev. Lett.}\ }\textbf
  {\bibinfo {volume} {109}},\ \bibinfo {pages} {266406} (\bibinfo {year}
  {2012})}\BibitemShut {NoStop}%
\bibitem [{\citenamefont {Latzke}\ \emph {et~al.}(2019)\citenamefont {Latzke},
  \citenamefont {Ojeda-Aristizabal}, \citenamefont {Denlinger}, \citenamefont
  {Reno}, \citenamefont {Zettl},\ and\ \citenamefont
  {Lanzara}}]{doi:10.1021/acsnano.9b04536}%
  \BibitemOpen
  \bibfield  {author} {\bibinfo {author} {\bibfnamefont {D.~W.}\ \bibnamefont
  {Latzke}}, \bibinfo {author} {\bibfnamefont {C.}~\bibnamefont
  {Ojeda-Aristizabal}}, \bibinfo {author} {\bibfnamefont {J.~D.}\ \bibnamefont
  {Denlinger}}, \bibinfo {author} {\bibfnamefont {R.}~\bibnamefont {Reno}},
  \bibinfo {author} {\bibfnamefont {A.}~\bibnamefont {Zettl}}, \ and\ \bibinfo
  {author} {\bibfnamefont {A.}~\bibnamefont {Lanzara}},\ }\bibfield  {title}
  {\enquote {\bibinfo {title} {Orbital character effects in the photon energy
  and polarization dependence of pure c60 photoemission},}\ }\href {\doibase
  10.1021/acsnano.9b04536} {\bibfield  {journal} {\bibinfo  {journal} {ACS
  Nano}\ }\textbf {\bibinfo {volume} {13}},\ \bibinfo {pages} {12710--12718}
  (\bibinfo {year} {2019})},\ \bibinfo {note} {pMID: 31638764},\ \Eprint
  {http://arxiv.org/abs/https://doi.org/10.1021/acsnano.9b04536}
  {https://doi.org/10.1021/acsnano.9b04536} \BibitemShut {NoStop}%
\bibitem [{\citenamefont {Wang}\ \emph {et~al.}(2012)\citenamefont {Wang},
  \citenamefont {Richard}, \citenamefont {Huang}, \citenamefont {Miao},
  \citenamefont {Cevey}, \citenamefont {Xu}, \citenamefont {Sun}, \citenamefont
  {Qian}, \citenamefont {Xu}, \citenamefont {Shi}, \citenamefont {Hu},
  \citenamefont {Dai},\ and\ \citenamefont {Ding}}]{PhysRevB.85.214518}%
  \BibitemOpen
  \bibfield  {author} {\bibinfo {author} {\bibfnamefont {X.-P.}\ \bibnamefont
  {Wang}}, \bibinfo {author} {\bibfnamefont {P.}~\bibnamefont {Richard}},
  \bibinfo {author} {\bibfnamefont {Y.-B.}\ \bibnamefont {Huang}}, \bibinfo
  {author} {\bibfnamefont {H.}~\bibnamefont {Miao}}, \bibinfo {author}
  {\bibfnamefont {L.}~\bibnamefont {Cevey}}, \bibinfo {author} {\bibfnamefont
  {N.}~\bibnamefont {Xu}}, \bibinfo {author} {\bibfnamefont {Y.-J.}\
  \bibnamefont {Sun}}, \bibinfo {author} {\bibfnamefont {T.}~\bibnamefont
  {Qian}}, \bibinfo {author} {\bibfnamefont {Y.-M.}\ \bibnamefont {Xu}},
  \bibinfo {author} {\bibfnamefont {M.}~\bibnamefont {Shi}}, \bibinfo {author}
  {\bibfnamefont {J.-P.}\ \bibnamefont {Hu}}, \bibinfo {author} {\bibfnamefont
  {X.}~\bibnamefont {Dai}}, \ and\ \bibinfo {author} {\bibfnamefont
  {H.}~\bibnamefont {Ding}},\ }\bibfield  {title} {\enquote {\bibinfo {title}
  {Orbital characters determined from fermi surface intensity patterns using
  angle-resolved photoemission spectroscopy},}\ }\href {\doibase
  10.1103/PhysRevB.85.214518} {\bibfield  {journal} {\bibinfo  {journal} {Phys.
  Rev. B}\ }\textbf {\bibinfo {volume} {85}},\ \bibinfo {pages} {214518}
  (\bibinfo {year} {2012})}\BibitemShut {NoStop}%
\bibitem [{\citenamefont {Volckaert}\ \emph {et~al.}(2019)\citenamefont
  {Volckaert}, \citenamefont {Rostami}, \citenamefont {Biswas}, \citenamefont
  {Markovi\ifmmode~\acute{c}\else \'{c}\fi{}}, \citenamefont {Andreatta},
  \citenamefont {Sanders}, \citenamefont {Majchrzak}, \citenamefont {Cacho},
  \citenamefont {Chapman}, \citenamefont {Wyatt}, \citenamefont {Springate},
  \citenamefont {Lizzit}, \citenamefont {Bignardi}, \citenamefont {Lizzit},
  \citenamefont {Mahatha}, \citenamefont {Bianchi}, \citenamefont {Lanata},
  \citenamefont {King}, \citenamefont {Miwa}, \citenamefont {Balatsky},
  \citenamefont {Hofmann},\ and\ \citenamefont
  {Ulstrup}}]{PhysRevB.100.241406}%
  \BibitemOpen
  \bibfield  {author} {\bibinfo {author} {\bibfnamefont {K.}~\bibnamefont
  {Volckaert}}, \bibinfo {author} {\bibfnamefont {H.}~\bibnamefont {Rostami}},
  \bibinfo {author} {\bibfnamefont {D.}~\bibnamefont {Biswas}}, \bibinfo
  {author} {\bibfnamefont {I.}~\bibnamefont {Markovi\ifmmode~\acute{c}\else
  \'{c}\fi{}}}, \bibinfo {author} {\bibfnamefont {F.}~\bibnamefont
  {Andreatta}}, \bibinfo {author} {\bibfnamefont {C.~E.}\ \bibnamefont
  {Sanders}}, \bibinfo {author} {\bibfnamefont {P.}~\bibnamefont {Majchrzak}},
  \bibinfo {author} {\bibfnamefont {C.}~\bibnamefont {Cacho}}, \bibinfo
  {author} {\bibfnamefont {R.~T.}\ \bibnamefont {Chapman}}, \bibinfo {author}
  {\bibfnamefont {A.}~\bibnamefont {Wyatt}}, \bibinfo {author} {\bibfnamefont
  {E.}~\bibnamefont {Springate}}, \bibinfo {author} {\bibfnamefont
  {D.}~\bibnamefont {Lizzit}}, \bibinfo {author} {\bibfnamefont
  {L.}~\bibnamefont {Bignardi}}, \bibinfo {author} {\bibfnamefont
  {S.}~\bibnamefont {Lizzit}}, \bibinfo {author} {\bibfnamefont {S.~K.}\
  \bibnamefont {Mahatha}}, \bibinfo {author} {\bibfnamefont {M.}~\bibnamefont
  {Bianchi}}, \bibinfo {author} {\bibfnamefont {N.}~\bibnamefont {Lanata}},
  \bibinfo {author} {\bibfnamefont {P.~D.~C.}\ \bibnamefont {King}}, \bibinfo
  {author} {\bibfnamefont {J.~A.}\ \bibnamefont {Miwa}}, \bibinfo {author}
  {\bibfnamefont {A.~V.}\ \bibnamefont {Balatsky}}, \bibinfo {author}
  {\bibfnamefont {P.}~\bibnamefont {Hofmann}}, \ and\ \bibinfo {author}
  {\bibfnamefont {S.}~\bibnamefont {Ulstrup}},\ }\bibfield  {title} {\enquote
  {\bibinfo {title} {Momentum-resolved linear dichroism in bilayer
  ${\mathrm{mos}}_{2}$},}\ }\href {\doibase 10.1103/PhysRevB.100.241406}
  {\bibfield  {journal} {\bibinfo  {journal} {Phys. Rev. B}\ }\textbf {\bibinfo
  {volume} {100}},\ \bibinfo {pages} {241406} (\bibinfo {year}
  {2019})}\BibitemShut {NoStop}%
\bibitem [{\citenamefont {Cao}\ \emph {et~al.}(2013)\citenamefont {Cao},
  \citenamefont {Waugh}, \citenamefont {Zhang}, \citenamefont {Luo},
  \citenamefont {Wang}, \citenamefont {Reber}, \citenamefont {Mo},
  \citenamefont {Xu}, \citenamefont {Yang}, \citenamefont {Schneeloch},
  \citenamefont {Gu}, \citenamefont {Brahlek}, \citenamefont {Bansal},
  \citenamefont {Oh}, \citenamefont {Zunger},\ and\ \citenamefont
  {Dessau}}]{Cao2013}%
  \BibitemOpen
  \bibfield  {author} {\bibinfo {author} {\bibfnamefont {Y.}~\bibnamefont
  {Cao}}, \bibinfo {author} {\bibfnamefont {J.~A.}\ \bibnamefont {Waugh}},
  \bibinfo {author} {\bibfnamefont {X.-W.}\ \bibnamefont {Zhang}}, \bibinfo
  {author} {\bibfnamefont {J.-W.}\ \bibnamefont {Luo}}, \bibinfo {author}
  {\bibfnamefont {Q.}~\bibnamefont {Wang}}, \bibinfo {author} {\bibfnamefont
  {T.~J.}\ \bibnamefont {Reber}}, \bibinfo {author} {\bibfnamefont {S.~K.}\
  \bibnamefont {Mo}}, \bibinfo {author} {\bibfnamefont {Z.}~\bibnamefont {Xu}},
  \bibinfo {author} {\bibfnamefont {A.}~\bibnamefont {Yang}}, \bibinfo {author}
  {\bibfnamefont {J.}~\bibnamefont {Schneeloch}}, \bibinfo {author}
  {\bibfnamefont {G.~D.}\ \bibnamefont {Gu}}, \bibinfo {author} {\bibfnamefont
  {M.}~\bibnamefont {Brahlek}}, \bibinfo {author} {\bibfnamefont
  {N.}~\bibnamefont {Bansal}}, \bibinfo {author} {\bibfnamefont
  {S.}~\bibnamefont {Oh}}, \bibinfo {author} {\bibfnamefont {A.}~\bibnamefont
  {Zunger}}, \ and\ \bibinfo {author} {\bibfnamefont {D.~S.}\ \bibnamefont
  {Dessau}},\ }\bibfield  {title} {\enquote {\bibinfo {title} {Mapping the
  orbital wavefunction of the surface states in three-dimensional topological
  insulators ${\mathrm{mos}}_{2}$},}\ }\href {\doibase 10.1038/nphys2685}
  {\bibfield  {journal} {\bibinfo  {journal} {Nat. Phys.}\ }\textbf {\bibinfo
  {volume} {9}},\ \bibinfo {pages} {499} (\bibinfo {year} {2013})}\BibitemShut
  {NoStop}%
\bibitem [{\citenamefont {Scholz}\ \emph {et~al.}(2013)\citenamefont {Scholz},
  \citenamefont {S\'anchez-Barriga}, \citenamefont {Braun}, \citenamefont
  {Marchenko}, \citenamefont {Varykhalov}, \citenamefont {Lindroos},
  \citenamefont {Wang}, \citenamefont {Lin}, \citenamefont {Bansil},
  \citenamefont {Min\'ar}, \citenamefont {Ebert}, \citenamefont {Volykhov},
  \citenamefont {Yashina},\ and\ \citenamefont {Rader}}]{Rader}%
  \BibitemOpen
  \bibfield  {author} {\bibinfo {author} {\bibfnamefont {M.~R.}\ \bibnamefont
  {Scholz}}, \bibinfo {author} {\bibfnamefont {J.}~\bibnamefont
  {S\'anchez-Barriga}}, \bibinfo {author} {\bibfnamefont {J.}~\bibnamefont
  {Braun}}, \bibinfo {author} {\bibfnamefont {D.}~\bibnamefont {Marchenko}},
  \bibinfo {author} {\bibfnamefont {A.}~\bibnamefont {Varykhalov}}, \bibinfo
  {author} {\bibfnamefont {M.}~\bibnamefont {Lindroos}}, \bibinfo {author}
  {\bibfnamefont {Y.~J.}\ \bibnamefont {Wang}}, \bibinfo {author}
  {\bibfnamefont {H.}~\bibnamefont {Lin}}, \bibinfo {author} {\bibfnamefont
  {A.}~\bibnamefont {Bansil}}, \bibinfo {author} {\bibfnamefont
  {J.}~\bibnamefont {Min\'ar}}, \bibinfo {author} {\bibfnamefont
  {H.}~\bibnamefont {Ebert}}, \bibinfo {author} {\bibfnamefont
  {A.}~\bibnamefont {Volykhov}}, \bibinfo {author} {\bibfnamefont {L.~V.}\
  \bibnamefont {Yashina}}, \ and\ \bibinfo {author} {\bibfnamefont
  {O.}~\bibnamefont {Rader}},\ }\bibfield  {title} {\enquote {\bibinfo {title}
  {Reversal of the circular dichroism in angle-resolved photoemission from
  ${\mathrm{bi}}_{2}{\mathrm{te}}_{3}$},}\ }\href {\doibase
  10.1103/PhysRevLett.110.216801} {\bibfield  {journal} {\bibinfo  {journal}
  {Phys. Rev. Lett.}\ }\textbf {\bibinfo {volume} {110}},\ \bibinfo {pages}
  {216801} (\bibinfo {year} {2013})}\BibitemShut {NoStop}%
\bibitem [{\citenamefont {Bigi}\ \emph {et~al.}(2021)\citenamefont {Bigi},
  \citenamefont {Mazzola}, \citenamefont {Fujii}, \citenamefont {Vobornik},
  \citenamefont {Panaccione},\ and\ \citenamefont {Rossi}}]{Bigi}%
  \BibitemOpen
  \bibfield  {author} {\bibinfo {author} {\bibfnamefont {C.}~\bibnamefont
  {Bigi}}, \bibinfo {author} {\bibfnamefont {F.}~\bibnamefont {Mazzola}},
  \bibinfo {author} {\bibfnamefont {J.}~\bibnamefont {Fujii}}, \bibinfo
  {author} {\bibfnamefont {I.}~\bibnamefont {Vobornik}}, \bibinfo {author}
  {\bibfnamefont {G.}~\bibnamefont {Panaccione}}, \ and\ \bibinfo {author}
  {\bibfnamefont {G.}~\bibnamefont {Rossi}},\ }\bibfield  {title} {\enquote
  {\bibinfo {title} {Measuring spin-polarized electronic states of quantum
  materials: $2h\ensuremath{-}{\mathrm{nbse}}_{2}$},}\ }\href {\doibase
  10.1103/PhysRevB.103.245142} {\bibfield  {journal} {\bibinfo  {journal}
  {Phys. Rev. B}\ }\textbf {\bibinfo {volume} {103}},\ \bibinfo {pages}
  {245142} (\bibinfo {year} {2021})}\BibitemShut {NoStop}%
\end{thebibliography}%

\newpage

\newpage
\section*{Supplementary Materials}
\section{Photon energy dependence of energy distribution curves compared to calculated photoionization cross sections}
\begin{figure} [H]
\includegraphics[width=8cm]{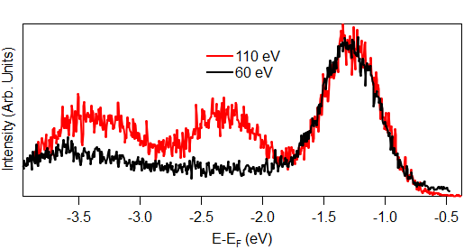}
\includegraphics[width=8cm]{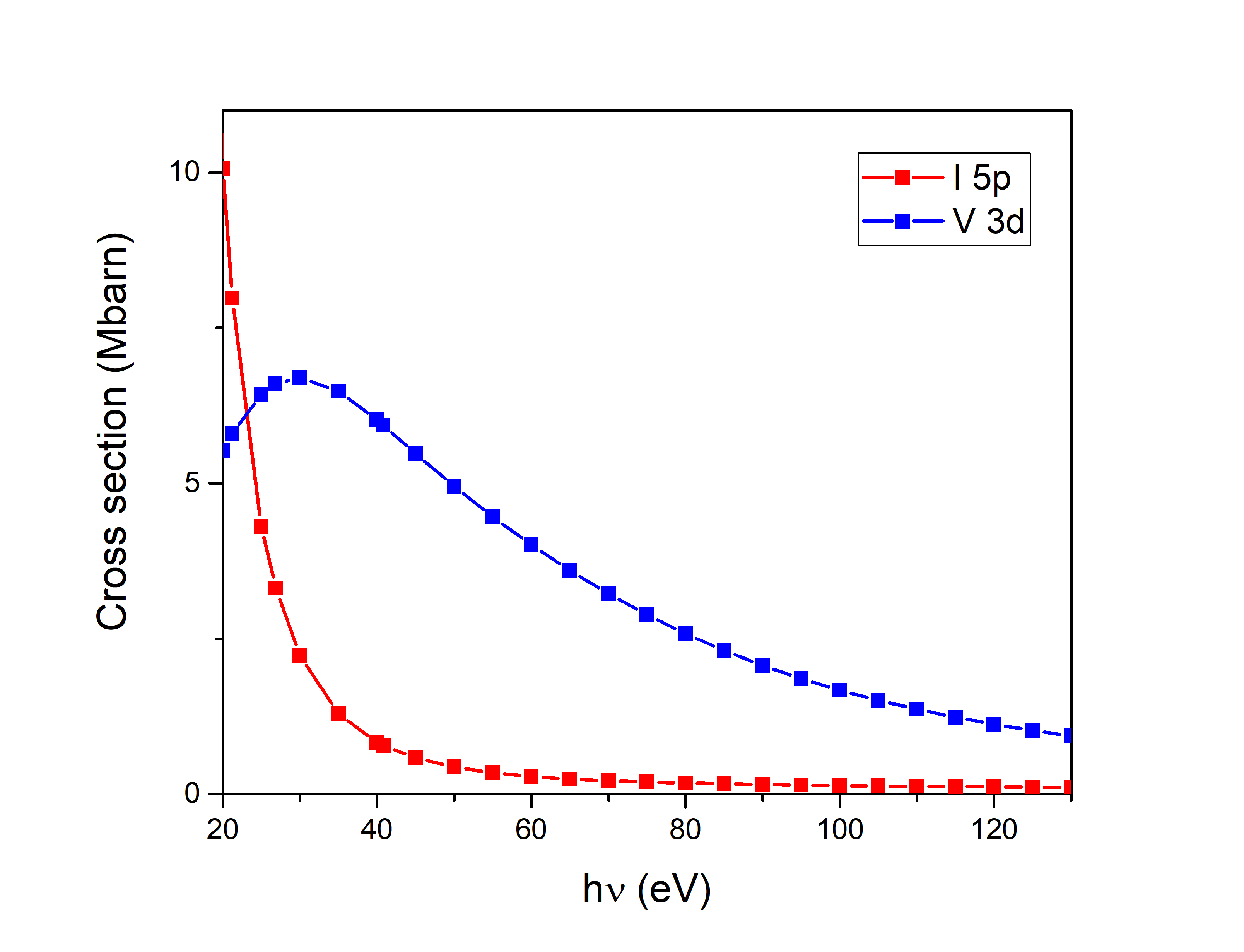}
\caption{\label{crosssection} \textbf{Supplementary materials} Left: Energy distribution curves at k$_y$=1.1\AA$^{-1}$ for photon energies h$\nu$= 60 eV (black) and 110 eV (red). Right: Calculated photoionization cross sections for V 3d (blue) and I 5p (red) related states from \onlinecite{YEH19851,CrossSectionParry}.} 
\end{figure}

The contrast of energy distribution curves at different photon energies (Figure \ref{crosssection}) with the calculated atomic subshell photoionization cross sections for the V 3d and I 5p related states as a function of photon energy \cite{YEH19851,CrossSectionParry}, supports our identification of the flat band near the top of the valence band as V 3d electron-derived. At h$\nu$ = 60 eV, the observed intensity at 1.2 eV below the Fermi energy (that we identify as V 3d derived states) is larger compared to the observed intensity at higher binding energies (below 1.5 eV) that we identify as I 5p derived states. This is consistent with the calculated cross sections for V 3d and I 5p related states at h$\nu$=60 eV. 
In the same way, at h$\nu$ = 110 eV, the observed intensity at 1.2 eV below the Fermi energy is comparable to the observed intensity at higher binding energies (below 1.5 eV) which is also consistent with the calculated cross sections for V 3d and I 5p related states at h$\nu$=110 eV.  

\end{document}